\documentclass[10pt]{article}

\usepackage[T1]{fontenc}
\usepackage{lmodern}
\usepackage[margin=1in]{geometry}
\usepackage{amsmath,amssymb}
\usepackage{booktabs,array}
\usepackage{graphicx}
\usepackage{algorithm}
\usepackage{algpseudocode}
\usepackage{xcolor}
\usepackage{listings}
\usepackage{tikz}
\usetikzlibrary{arrows.meta,positioning}
\usepackage[round]{natbib}
\usepackage{microtype}
\usepackage[colorlinks=true,linkcolor=blue,citecolor=blue,urlcolor=blue]{hyperref}

\hypersetup{
  pdftitle={GC-MoE: Overlapping Expert-Parallel Token Dispatch with FP8/FP4 Grouped GEMM via CUDA Green Contexts},
  pdfauthor={Yong Li, Kan Liu, Tao Lan},
  pdfsubject={An open-source MoE inference operator that uses CUDA Green Contexts to overlap expert-parallel token dispatch with FP8/FP4 grouped GEMM on NVIDIA Hopper and Blackwell GPUs},
  pdfkeywords={mixture of experts, expert parallelism, CUDA Green Contexts, FP8, FP4, grouped GEMM}
}

\graphicspath{{figures/}}

\newcolumntype{L}[1]{>{\raggedright\arraybackslash}p{#1}}

\newcommand{\best}[1]{\textbf{#1}}

\title{GC-MoE: Overlapping Expert-Parallel Token Dispatch with\\
FP8/FP4 Grouped GEMM via CUDA Green Contexts}

\author{Yong Li, Kan Liu, and Tao Lan\\
\small \texttt{\{brucelee.ly, liukan.lk, tao.lant\}@alibaba-inc.com}\\
\small \url{https://github.com/Bruce-Lee-LY/gc-moe}}

\date{November 2, 2026}

\begin{document}

\maketitle

\begin{abstract}
Mixture-of-experts (MoE) models are standard for large language model (LLM) serving, and expert parallelism (EP) makes all-to-all token dispatch, grouped GEMMs, and the combine step recurrent costs in every layer. In decode, each rank typically holds only 1--256 tokens, so dispatch latency and weight traffic dominate while tensor-core utilization is inherently low; because the grouped GEMM is bandwidth-bound, a decode layer does not need all SMs for arithmetic. In state-of-the-art fused operators such as DeepGEMM's Mega MoE, dispatch and computation overlap only by sharing every SM: a few dispatch warps per SM compete with the TMA/MMA and epilogue warps for registers, shared memory, and issue slots, every phase is bound to one CTA's combined resource budget, the synchronization domain spans the entire SM array, and every replay ends with a workspace-clean pass and a cross-rank tag-2 barrier. This paper presents GC-MoE, an open-source FP8/FP4 MoE inference operator for NVIDIA Hopper and Blackwell GPUs that uses CUDA Green Contexts to split the device at the driver level into a small communication partition and a large compute partition. The split gives each role its own kernel, resource budget, and synchronization domain, so dispatch and GEMM can be tuned independently. A three-node CUDA Graph (four nodes on Hopper with fused shared experts) launches the dispatch and grouped-GEMM kernels as concurrent roots on the two partitions; a token-ring handshake lets computation start on the first filled ring blocks, and double-buffered receive counters remove the trailing cross-rank barrier between replays. The compute partition runs operand-swapped tcgen05 block-scaled MMA on Blackwell and register-A WGMMA with in-register FP4 decoding on Hopper, with fused SwiGLU, requantization, and optional fused shared experts. On eight H20 GPUs GC-MoE accelerates DeepSeek-V4.1-Flash decode by 1.188$\times$ on average on both the routed and shared-expert suites (peaks 1.296$\times$ and 1.353$\times$, respectively); on eight B300 GPUs it accelerates GLM-5.2 and DeepSeek-V4.1-Flash by 1.106$\times$ on average without shared experts and 1.123$\times$ with fused shared experts (peak 1.209$\times$), outperforming DeepGEMM's Mega MoE at all 54 measured points at matched numerical accuracy (the nine H20 shared-expert points compare against an unfused shared-expert baseline). GC-MoE reserves only 20 of 148 SMs on B300 for dispatch.
\end{abstract}

\section{Introduction}
\label{sec:intro}

Mixture-of-experts (MoE) architectures have become the dominant way to scale large language models without proportionally scaling inference cost~\citep{shazeer2017,gshard,switch,mixtral,deepseekv2,deepseekv3}. In production serving, MoE layers are sharded with expert parallelism (EP): each GPU owns a subset of experts, and every layer routes each token to its top-$k$ experts through an all-to-all token dispatch, computes the routed experts locally, and combines the partial outputs back to the source ranks. Decode dominates the serving cost of interactive applications, and in decode each rank typically processes only a handful to a few hundred tokens per layer. Under such small batch sizes, the MoE layer is not tensor-core bound: its cost is dominated by cross-rank token movement and by streaming quantized expert weights from HBM.

Because of this cost structure, kernel-level MoE optimization has converged on fused, persistent operators. DeepGEMM's \emph{Mega MoE} is the state of the art on NVIDIA Blackwell (and, via a reference SM90 port, on Hopper): a single fused kernel performs the expert-parallel dispatch, the FP8$\times$FP4 grouped GEMMs with SwiGLU fused into the first epilogue, and the combine, with warp specialization inside one kernel and block-scaled tensor-core instructions~\citep{deepgemm,deepseekv3}. It already overlaps dispatch with computation, but only by sharing every SM (streaming multiprocessor): four dispatch warps per SM pull tokens into a per-block ring while two TMA-load warps (TMA: tensor-memory-accelerator copy engine), one MMA-issue warp (MMA: matrix multiply--accumulate), a scheduler warp, and eight epilogue warps consume ring blocks, run the GEMM, and perform the combine. The compute warps therefore never own the whole SM---the dispatch warps and their staging buffers are carved out of each SM's register file and shared memory---and every replay ends with a workspace-clean pass followed by a cross-rank tag-2 barrier. The clean overlaps the combine, but the barrier is still a cross-rank round trip paid on every replay. The kernel's synchronization domain is the whole device: its prologue and tail grid syncs include every SM of the device (148 on B300), so each phase waits for the slowest CTA (cooperative thread array), and the same CTA budget must host all phases, so a local change to one phase is a change to the whole kernel. These effects inflate decode latency at 1--256 tokens per rank.

Two further trends sharpen the problem. First, low-precision formats have moved from FP8 activations to FP4 weights: weight traffic, not arithmetic, sets the decode lower bound, and the two GPU generations expose different instruction sets for mixed FP8$\times$FP4 computation (tcgen05 block-scaled MMA on Blackwell, WGMMA with software FP4 decode on Hopper). Second, existing MoE systems optimize communication--computation overlap \emph{across} micro-batches, layers, or experts~\citep{lina,fastermoe}, or rely on stream-level concurrency, but they do not isolate SM resources \emph{inside} the operator: a conventional kernel launched on the full device occupies all SMs, and concurrent kernels on separate streams compete for the same SM array with no hardware guarantee of co-residency or performance isolation. CUDA Green Contexts, introduced in CUDA 12.4, provide driver-level SM partitioning with independent workqueues, i.e., exactly the primitive needed to co-schedule a communication kernel and a compute kernel deterministically. Prior systems have used green contexts for \emph{isolation} between inference processes, co-located models, or prefill/decode requests~\citep{gcisolation,eclip,agentserve}; to our knowledge, none uses them to overlap the communication and computation phases \emph{within} a single MoE operator. Deeper fusion alone does not address the decode bottleneck: because the grouped GEMM is bandwidth-bound, extra SMs buy little arithmetic, while a fused kernel couples the phases' resource and synchronization budgets.

This paper presents GC-MoE, an open-source MoE inference operator for Hopper (SM90) and Blackwell (SM103) GPUs that reorganizes the MoE layer as a three-node CUDA Graph (four nodes on Hopper with fused shared experts) across two CUDA Green Contexts. A small \emph{communication partition} (20 of 148 SMs on B300) runs the dispatch kernel; a large \emph{compute partition} runs the grouped-GEMM kernel concurrently. The two kernels are independent graph roots, so the driver schedules them simultaneously on disjoint SM sets. A token-ring protocol with generation-tagged counters lets the GEMM kernel begin consuming a block of dispatched tokens as soon as that block is filled, without waiting for dispatch to finish. A finalize kernel then aggregates the top-$k$ and shared-expert partials into the output. Double-buffered receive counters and device-persistent synchronization state remove the trailing cross-rank barrier, so consecutive CUDA Graph replays need no cross-rank rendezvous between them. Because the grouped GEMM is bandwidth-bound at decode, ceding 20 SMs to dispatch costs little compute throughput; because each role is an ordinary kernel with its own thread, register, shared-memory, and synchronization budget, the dispatch and GEMM paths can also be tuned independently. The compute kernel implements FP8-activation$\times$FP4-weight grouped GEMM for both architectures: an operand-swapped tcgen05 block-scaled MMA path on Blackwell that services FP4 routed and FP8 shared weights with one instruction sequence, and a register-A WGMMA path on Hopper with in-register FP4$\rightarrow$E4M3 decoding and a window-based scale repacking. SwiGLU, routing-weight application, and FP8 requantization are fused into the first GEMM epilogue; shared experts can be fused into the same task stream.

Our contributions are:
\begin{itemize}
\item \textbf{Decode-aware green-context decomposition.} To our knowledge, GC-MoE is the first MoE operator that uses CUDA Green Contexts to give token dispatch and expert computation disjoint, driver-aligned SM partitions, so dispatch runs concurrently with the grouped GEMM on dedicated SMs instead of sharing every SM; balanced workqueues avoid implicit serialization between the two contexts. At decode the GEMM is bandwidth-bound, so the 20-SM dispatch partition costs little compute throughput, and each role is an ordinary kernel with its own resource budget and synchronization domain, which allows the dispatch and GEMM paths to be optimized independently (Sections~\ref{sec:motivation} and~\ref{sec:partition}).
\item \textbf{Fine-grained cross-partition synchronization.} A token ring with generation-tagged full/empty counters and a 64-bit W1-completion bitmap carries the baseline's block-granular handshake across the partition boundary, so the compute kernel starts on the first filled ring block; double-buffered receive counters eliminate the trailing cross-rank barrier between graph replays; all synchronization state survives replay without host intervention (Sections~\ref{sec:overlap} and~\ref{sec:combine}).
\item \textbf{Dual-architecture FP8$\times$FP4 grouped GEMM.} Operand-swapped tcgen05 block-scaled MMA with K128 tile-prepacked FP4 weights on Blackwell, and register-A WGMMA with in-register FP4 decoding and exact window-based scale repacking on Hopper, both with fused SwiGLU, requantization, and optional fused shared experts (Sections~\ref{sec:gemm} and~\ref{sec:se}).
\item \textbf{Systematic evaluation.} On 8$\times$H20 and 8$\times$B300, GC-MoE outperforms DeepGEMM's Mega MoE at all 54 measured decode points: 1.188$\times$ mean on H20 (DeepSeek-V4.1-Flash, peaks 1.296$\times$ routed and 1.353$\times$ with shared experts), 1.106$\times$ mean on B300 without shared experts and 1.123$\times$ with fused shared experts (GLM-5.2 and DeepSeek-V4.1-Flash, peak 1.209$\times$). Accuracy is matched: cosine $\geq 0.97$ against BF16 everywhere and minimum cosine $\geq 0.9999997$ against the baseline on both H20 suites; the nine H20 shared-expert points compare against an unfused shared-expert baseline (Section~\ref{sec:se-eval}). The operator is open source.
\end{itemize}

\section{Background}
\label{sec:background}

\paragraph{Notation.} $P$ denotes the number of ranks (GPUs), $E$ the total number of routed experts, $e=E/P$ the local experts per rank, $H$ the hidden size, $I$ the intermediate size, $T$ the number of tokens per rank, and $k$ the top-$k$ routing factor. $B_M$ is the token block size of the grouped GEMM; $B_N=128$ and $B_K=128$ are the fixed MMA tile dimensions on the Blackwell path. FP8 refers to E4M3, FP4 to E2M1 packed two values per byte, and UE8M0 to the unsigned power-of-two exponent scale used by the microscaling recipes.

\subsection{MoE layers under expert parallelism}
\label{sec:moe}

A routed-expert MoE layer computes, for token $t$,
\begin{equation}
y_t=\sum_{j\in\mathrm{TopK}(t)} g_{t,j}\,W_2^{(j)}\,\mathrm{SwiGLU}\!\left(W_1^{(j)}x_t\right)
\;+\;\underbrace{W_2^{s}\,\mathrm{SwiGLU}\!\left(W_1^{s}x_t\right)}_{\text{optional shared expert}},
\label{eq:moe}
\end{equation}
where $W_1^{(j)}\in\mathbb{R}^{2I\times H}$ stacks the gate and up projections, $W_2^{(j)}\in\mathbb{R}^{H\times I}$ is the down projection, and $g_{t,j}$ is the routing weight. Under expert parallelism the $E$ experts are distributed evenly, $e=E/P$ per rank. Executing one layer therefore decomposes into four stages: (i)~\emph{dispatch}, an all-to-all exchange that moves each token, its activation scales, and routing metadata to the ranks that own its selected experts; (ii)~\emph{Grouped GEMM 1} over the per-expert token groups, with SwiGLU in the epilogue; (iii)~\emph{Grouped GEMM 2} producing the down projections; and (iv)~\emph{combine}, the weighted reduction of the $k$ partial outputs back to the source ranks. On the Blackwell path the routed activations are FP8 with one UE8M0 scale per 32 values along the contraction dimension, the routed weights are FP4 at the same 32-value scale granularity, and the shared-expert weights are FP8; the Hopper path instead consumes E4M3FN activations with contiguous FP32 K128 scales (Section~\ref{sec:formats} and Table~\ref{tab:paths}).

In decode at small batch, each rank owns 1--256 tokens, so each local expert receives a small number of tokens and the per-expert GEMM has a tiny M dimension (often 16--32 rows after grouping). For GLM-5.2's 256-expert, 8-rank, top-8 configuration, balanced routing gives $T\cdot k\cdot P/E=T/4$ tokens per expert; at one token per rank an expert receives 0.25 tokens on average, and the padded M tile is almost entirely padding. The grouped GEMM is then bandwidth-bound: its cost is dominated by streaming the local expert weights (FP4) and scale metadata from HBM rather than by arithmetic. At the same time, the dispatch stage is latency-bound: its critical path is the cross-rank exchange plus peer-atomic accounting, and it needs very few SMs to saturate NVLink. Because a bandwidth-bound GEMM gains little from extra SMs, the layer does not need the whole device for arithmetic; yet a single fused kernel spans all SMs, and a device-wide synchronization domain is expensive when the layer itself is short. These observations---dispatch needs SMs but no tensor cores, decode GEMM needs tensor cores but is bandwidth-bound, and all-SM synchronization is costly at decode---are the premise of this work.

\subsection{GPU execution model: Hopper, Blackwell, Green Contexts, and graphs}
\label{sec:execution}

\paragraph{Hopper (SM90).} On H20, kernels target \texttt{sm\_90a}. The architecture provides warpgroup MMA (WGMMA), asynchronous TMA copies, and warp specialization; a warpgroup of four warps issues \texttt{wgmma.mma\_async} instructions whose operands are read from shared memory or registers. There is no native FP4 MMA: FP4 weights must be decoded to E4M3 in software before entering the tensor core.

\paragraph{Blackwell (SM103).} On B300 (compute capability 10.3, an SM100-family part with 148 SMs), kernels target \texttt{sm\_103f}. The fifth-generation tensor core is programmed through tcgen05 instructions: MMA accumulates into tensor memory (TMEM), a two-CTA cluster pair can issue a single MMA cooperatively (\texttt{cta\_group::2}), and the block-scaled variant \texttt{kind::mxf8f6f4.}\allowbreak\texttt{block\_scale} multiplies mixed FP8/FP4 operands per 32-value group using hardware scale factors. MMA completion is signaled through an mbarrier; TMEM is then read back with \texttt{tcgen05.ld}, so the epilogue can be software-pipelined against the next MMA without a device-wide barrier.

\paragraph{CUDA Green Contexts.} A green context is a lightweight execution context that owns a subset of the device's streaming multiprocessors. The driver API exposes the device's SM resource (\texttt{cuDevice}\allowbreak\texttt{GetDevResource}), splits it by count (\texttt{cuDevSmResourceSplitByCount}), and creates a green context over the resulting resource (\texttt{cuGreenCtxCreate}). Kernels launched into a green context's stream are scheduled only on its SMs; two green contexts therefore run concurrently on disjoint SM sets with no MPS daemon and no process-level configuration. The split granularity and alignment are driver properties (minimum partition size and co-scheduling alignment), so the application requests a partition size and the driver determines the usable partition. Green contexts can additionally carry a workqueue configuration; the \emph{balanced} sharing scope with a concurrency limit of one gives each context its own hardware workqueue, avoiding implicit serialization when two contexts' kernels contend for one queue.

\paragraph{CUDA Graphs.} A CUDA Graph captures a kernel-node DAG once and replays it with negligible host overhead. Independent nodes without dependency edges are eligible to execute concurrently; node parameters are fixed at capture time and can be passed by value, which allows a graph to carry large descriptor tables (tensor maps, peer address tables) without per-launch pointer chasing. We use these properties to express the operator as a three-node child graph that is attached to an outer capture (Section~\ref{sec:combine}).

\subsection{Low-precision formats}
\label{sec:formats}

On the Blackwell path the operator consumes activations quantized by the caller to FP8 E4M3 with one UE8M0 scale per 32 contiguous values along $K$; four exponent bytes are packed into one \texttt{int32}. Routed weights are FP4 (E2M1) with the same K32 microscaling granularity, packed two values per byte; the shared-expert weights are FP8. This is the $(1,1,32)$ recipe: one scale per $1\times32$ value block for both operands. For shared experts we additionally support a $(1,128,128)$ recipe whose weight scales cover $128\times128$ blocks; in that case the caller expands block-row scales to one packed row per output row before weight transformation. FP4 formats and their scaling conventions are shared with NVFP4/MXFP4-style training and inference work~\citep{nvfp4,mxfp4}; on Hopper, activations use contiguous FP32 K128 scales and routed weights follow the MXFP4 convention (E2M1 with UE8M0 K32 scales). Weight transformation only rearranges already-quantized data and scales; it never quantizes floating-point weights.

\section{Motivation: the costs of monolithic fusion at decode}
\label{sec:motivation}

\begin{figure}[t]
\centering
\begin{tikzpicture}[font=\footnotesize, >={Stealth[length=1.8mm]},
  lane/.style={rounded corners=2pt, draw=black!25, line width=0.5pt},
  ph/.style={rounded corners=2pt, line width=0.5pt, draw=white},
  brk/.style={draw=black!45, line width=0.5pt}]
\begin{scope}[yshift=0.30cm]
\node[anchor=west, font=\small\bfseries, text=black!85] at (-0.3,9.18)
  {(a) DeepGEMM Mega MoE (baseline): one fused kernel, warp roles share every SM};
\node[anchor=east, font=\footnotesize, text=black!70, align=right] at (2.72,8.38)
  {all 148 SMs\\[-1pt]{\scriptsize (shared)}};
\node[lane, fill=white, minimum width=12.0cm, minimum height=0.86cm, anchor=west] at (3.0,8.38) {};
\fill[blue!35, ph] (3.0,7.95) rectangle (4.1,8.81);
\node[font=\tiny, align=center] at (3.55,8.38) {counts +\\tag-1};
\fill[blue!4, ph] (4.1,7.95) rectangle (14.9,8.81);
\fill[orange!28, ph] (4.1,8.42) rectangle (11.9,8.77);
\node[font=\scriptsize] at (7.8,8.595) {dispatch warps (4 per SM): pull};
\fill[red!14, ph] (11.9,8.42) rectangle (13.9,8.77);
\node[font=\tiny, align=center, text=red!45!black] at (12.9,8.595) {clean};
\fill[red!26, ph] (13.9,8.42) rectangle (14.9,8.77);
\node[font=\tiny, align=center, text=red!45!black] at (14.4,8.595) {tag-2};
\fill[teal!26, ph] (4.75,7.99) rectangle (11.9,8.34);
\node[font=\scriptsize] at (8.1,8.165) {TMA/MMA + epilogue warps: GEMM, scatter};
\fill[teal!17, ph] (11.9,7.99) rectangle (14.9,8.34);
\node[font=\scriptsize, text=teal!35!black] at (13.4,8.165) {epilogue: combine};
\draw[dashed, black!55, line width=0.4pt] (4.75,7.95) -- (4.75,7.89);
\node[anchor=north west, font=\tiny, text=black!60] at (4.80,7.88) {first ring block};
\draw[brk] (4.1,7.48) -- (11.9,7.48);
\draw[brk] (4.1,7.40) -- (4.1,7.56);
\draw[brk] (11.9,7.40) -- (11.9,7.56);
\node[font=\scriptsize, text=black!60, anchor=north, align=center] at (8.0,7.36)
  {dispatch warps and their staging buffers are carved out of\\every SM's register file and shared memory};
\draw[brk, <->] (3.0,6.62) -- (14.9,6.62);
\draw[brk] (3.0,6.53) -- (3.0,6.71);
\draw[brk] (14.9,6.53) -- (14.9,6.71);
\node[font=\footnotesize, text=black!75, anchor=north, align=center] at (8.95,6.50)
  {$T_{\mathrm{mega}} = t_{\mathrm{pro}} + \max(T^{\mathrm{shared}}_{\mathrm{pull}},\, T^{\mathrm{shared}}_{\mathrm{gemm}}) + \max(T_{\mathrm{combine}},\, t_{\mathrm{clean}} + t_{\mathrm{tag2}})$\\[1pt]
   {\scriptsize the clean overlaps the combine; the tag-2 barrier follows the clean}};
\node[anchor=west, font=\scriptsize, text=black!55] at (14.96,6.62) {$t$};
\end{scope}
\node[anchor=west, font=\small\bfseries, text=black!85] at (-0.3,5.62)
  {(b) GC-MoE: disjoint SM sets, dispatch off the compute SMs, no tag-2 barrier};
\node[anchor=east, font=\footnotesize, text=black!70, align=right] at (2.72,4.88)
  {20 SMs\\[-1pt]{\scriptsize comm}};
\node[lane, fill=white, minimum width=12.0cm, minimum height=0.66cm, anchor=west] at (3.0,4.88) {};
\fill[blue!35, ph] (3.0,4.55) rectangle (4.1,5.21);
\node[font=\tiny, align=center] at (3.55,4.88) {counts +\\tag-1};
\fill[orange!30, ph] (4.1,4.55) rectangle (11.4,5.21);
\node[font=\scriptsize] at (7.75,4.88) {role1: pull tokens, scales, and metadata};
\foreach \x in {4.55,5.05,5.55,6.05,6.55,7.05,7.55,8.05,8.55,9.05,9.55,10.05,10.55}
  {\draw[white, line width=1pt] (\x,4.55) -- (\x,4.74);}
\node[draw=black!35, dashed, rounded corners=2pt, fill=gray!8,
      minimum width=1.05cm, minimum height=0.66cm, anchor=west] at (12.2,4.88) {};
\node[font=\tiny, text=black!55, align=center] at (12.725,4.88) {replay\\$N{+}1$};
\node[anchor=east, font=\footnotesize, text=black!70, align=right] at (2.72,3.64)
  {128 SMs\\[-1pt]{\scriptsize compute}};
\node[lane, fill=white, minimum width=12.0cm, minimum height=0.66cm, anchor=west] at (3.0,3.64) {};
\fill[teal!26, ph] (4.75,3.31) rectangle (11.4,3.97);
\node[font=\scriptsize, align=center] at (8.08,3.64)
  {\textbf{role2}: grouped GEMM, no dispatch warps\\consumes ring blocks as they land};
\fill[orange!30, ph] (11.4,3.31) rectangle (12.2,3.97);
\node[anchor=west, font=\scriptsize, text=black!60] at (11.48,4.16) {role3: combine};
\draw[dashed, black!55, ->] (4.75,4.55) -- (4.75,3.97);
\node[anchor=west, font=\tiny, text=black!60] at (4.82,4.26) {first ring block};
\fill[green!12] (12.2,3.31) rectangle (14.9,3.97);
\draw[dashed, green!40!black!50] (12.2,3.31) rectangle (14.9,3.97);
\node[font=\scriptsize, text=green!30!black] at (13.55,3.64) {saved (illustrative)};
\draw[brk, <->] (3.0,2.95) -- (12.2,2.95);
\draw[brk] (3.0,2.86) -- (3.0,3.04);
\draw[brk] (12.2,2.86) -- (12.2,3.04);
\node[font=\footnotesize, text=black!75, anchor=north, align=center] at (7.6,2.83)
  {$T_{\mathrm{gc}} = t_{\mathrm{pro}} + \max(T^{\mathrm{ded}}_{\mathrm{pull}},\, T_{\mathrm{gemm}}) + T_{\mathrm{combine}}$\\[1pt]
   {\scriptsize no tag-2 barrier: parity counters and role3 bookkeeping replace the cross-rank round trip}};
\node[anchor=west, font=\scriptsize, text=black!55] at (12.26,2.95) {$t$};
\end{tikzpicture}
\caption{Execution timelines. (a)~On B300 the baseline is DeepGEMM's Mega MoE, one persistent fused kernel on all 148 SMs: after routing counts and a tag-1 cross-rank barrier, the dispatch warps stream tokens, scales, and routing metadata into a per-block token ring while the GEMM consumes each ring block as soon as it is full. The phases overlap, but on the same SMs, with the dispatch warps and their staging buffers carved out of every SM's register file and shared memory; the tail is $\max(T_{\mathrm{combine}}, t_{\mathrm{clean}}+t_{\mathrm{tag2}})$---the workspace clean overlaps the combine, but the tag-2 barrier is still a cross-rank round trip on every replay. (b)~GC-MoE runs role1 (dispatch) on the 20-SM communication context and role2 (grouped GEMM) plus role3 (combine) on the 128-SM compute context. Role2 starts when the first ring block is filled (dashed trigger), and parity-double-buffered receive counters remove the tag-2 barrier, so the next replay can start as soon as role3 completes (green band). Each role is an ordinary kernel with its own register, shared-memory, and synchronization budget, so the dispatch and GEMM paths can be optimized independently. Bars are illustrative; measured end-to-end speedups span 1.03--1.30$\times$ across the routed-expert sweeps and 1.01--1.35$\times$ including shared experts (Section~\ref{sec:results}). Partitions are the B300 values.}
\label{fig:timeline}
\end{figure}
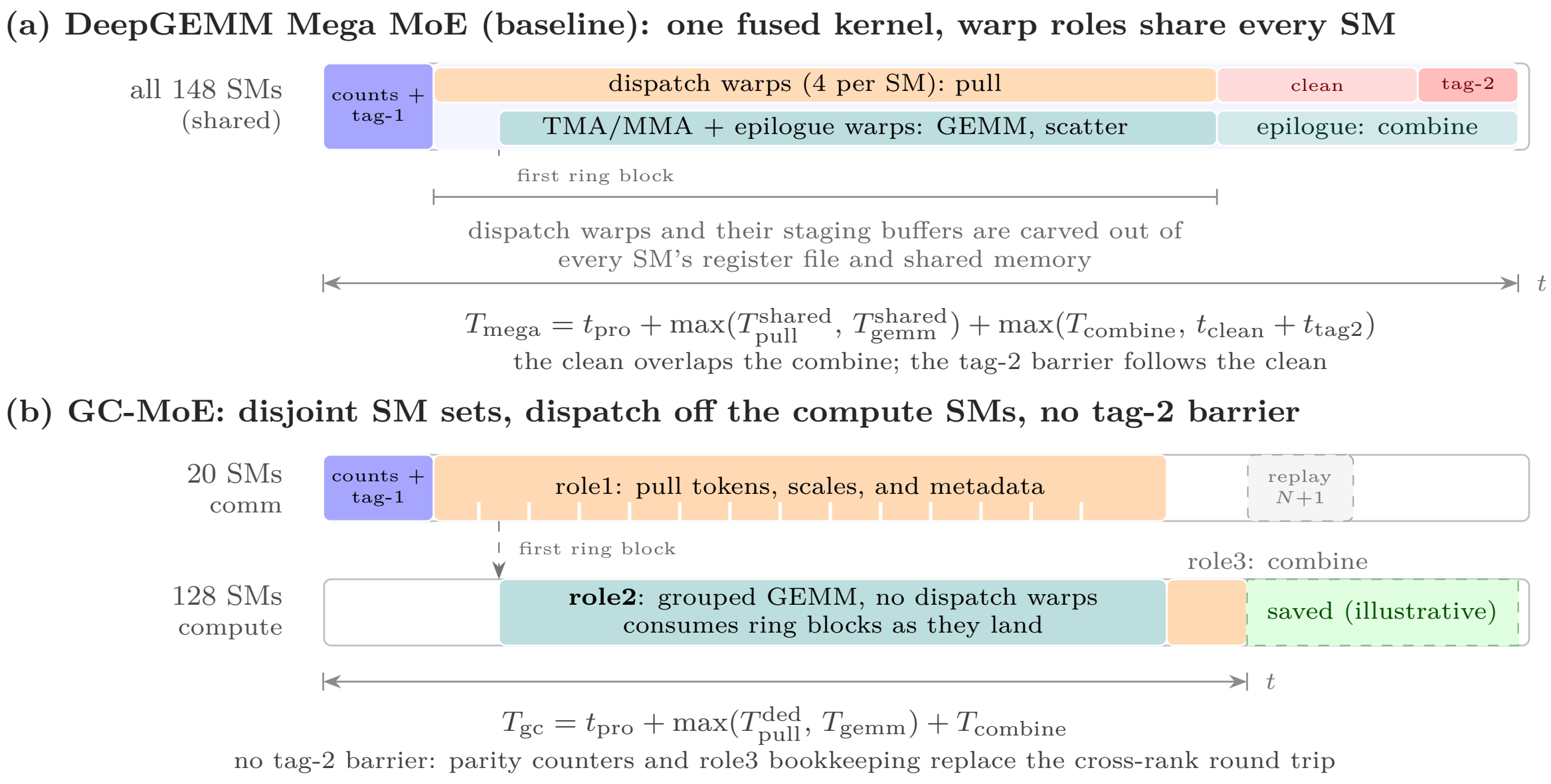

Figure~\ref{fig:timeline}(a) shows how DeepGEMM's Mega MoE executes a decode layer on B300. One persistent fused kernel occupies all 148 SMs. Warp specialization assigns each CTA four dispatch warps (128 threads), four non-epilogue warps (two TMA-load warps, one MMA-issue warp, and one scheduler warp; the last two run only in the leader CTA of each two-CTA cluster), and eight epilogue warps (256 threads) that also perform the combine; the kernel opens with routing accounting and a tag-1 cross-rank barrier. It then streams: the dispatch warps pull tokens, scales, and routing metadata into a per-block token ring, and the GEMM consumes a ring block as soon as that block is full. Dispatch and computation therefore \emph{do} overlap---but on the same SMs. The overlap is spatial sharing: the dispatch warps and their staging buffers are carved out of the register file and shared memory of every SM, the pull is issued by only four warps per SM, and the kernel closes with workspace cleaning and a tag-2 cross-rank barrier that every rank must pass before the next replay can reuse the workspace. The clean is explicitly overlapped with the combine reduction, but the tag-2 barrier follows the clean and is a cross-rank round trip. The critical path is
\begin{equation}
T_{\mathrm{mega}} = t_{\mathrm{pro}} + \max\!\left(T^{\mathrm{shared}}_{\mathrm{pull}},\, T^{\mathrm{shared}}_{\mathrm{gemm}}\right) + \max\!\left(T_{\mathrm{combine}},\, t_{\mathrm{clean}} + t_{\mathrm{tag2}}\right) ,
\label{eq:mega}
\end{equation}
where $t_{\mathrm{pro}}$ is the prologue, the two middle terms execute concurrently on one shared SM array, and the tail is the longer of the final combine and the workspace clean plus the tag-2 barrier.

Three costs follow. First, the compute warps never own the whole SM, so per-SM GEMM throughput is below what a dedicated SM delivers. Second, the synchronization domain is the entire device: the prologue and tail grid syncs include all 148 SMs, so each phase waits for the slowest CTA, and the tag-2 barrier adds a cross-rank round trip on every replay that the combine hides only when the combine itself outlasts the clean plus the barrier. Third, fusion couples the resource budgets: all roles share one CTA's 64\,512 registers and 227\,KB of shared memory, and even an idle role keeps its share reserved, so a deeper weight pipeline, a larger epilogue, or a cheaper dispatch path must fit the combined budget and preserve every other role's barriers and buffers; a local optimization is therefore paid for by the whole kernel.

GC-MoE removes all three. Role1 runs the dispatch on a 20-SM communication context that owns the whole SM budget for the pull, role2 runs the grouped GEMM on a 128-SM compute context that carries no dispatch warps, and role3's finalize grid is smaller still, so the grid syncs around the cross-rank tag-1 barrier span only the 20 dispatch SMs and role3's grid sync spans only its finalize grid, rather than all 148 SMs. Parity-double-buffered receive counters let role3 flip generations instead of waiting on a tag-2 barrier. Because the roles are now separate kernels, each has its own thread, register, shared-memory, and synchronization budget: the dispatch path can be tuned for latency and the GEMM path for streaming and MMA throughput without rebalancing a shared CTA, and each role can adopt phase-specific optimizations independently. GC-MoE additionally improves the GEMM itself by prepacking FP4 weights into K128 tiles (Section~\ref{sec:gemm}). In the measured H20 configuration, GC-MoE cuts end-to-end latency by 20.9--22.8\% (1.26--1.30$\times$ speedup) at 4--32 tokens per rank, the region where the grouped GEMM is still issue-bound at small $M$ and the shared-SM penalty and the tag-2 barrier weigh most; at 256 tokens per rank weight streaming dominates the layer, both operators approach their streaming limit, and the relative gain shrinks to the roughly fixed shared-SM and tail savings (Section~\ref{sec:results}).

Three straw-man alternatives do not solve the problem. (i)~\emph{Multiple CUDA streams without isolation}: a conventional kernel launched over the whole device fills every SM; a concurrently launched communication kernel must wait or steal SMs from the compute kernel, with no co-residency guarantee and no isolation from occupancy interference. (ii)~\emph{One fused kernel with warp-specialized roles} (the baseline): dispatch and compute can overlap, but only by sharing every SM, because a kernel cannot repartition SMs between roles without draining; the dispatch warps and their staging buffers are carved out of each SM's budget, the pull is issued by a handful of warps, and a trailing cross-rank barrier is still needed between invocations to protect workspace reuse. Every local change is also constrained by the whole kernel: the CTA's register and shared-memory budgets are already committed, and all roles' barriers and buffers must remain valid, so optimizing one phase in isolation is difficult. (iii)~\emph{Software-stage overlap across micro-batches or layers}: this hides communication with \emph{other} work but does not reduce a single layer's critical path, and it changes numerical scheduling of the model.

The design principles of GC-MoE follow directly. First, partition the SM array \emph{statically but natively}, using the driver's green-context SM split, so that dispatch and compute kernels are co-resident on disjoint SMs with hardware-level isolation. Second, realize every role as an ordinary kernel with its own thread, register, shared-memory, and synchronization budget, so that dispatch tuning and GEMM tuning are independent; the fused baseline instead charges every local change against one CTA's combined budget. Third, carry the baseline's block-granular token-ring handshake across the partition boundary, so the compute kernel starts on the first filled ring block rather than at the end of dispatch. Fourth, keep \emph{all synchronization state on device and persistent across graph replays}, and double-buffer cross-rank counters so that no trailing barrier is needed between replays. Fifth, specialize the compute kernel per architecture, because the instruction sets for mixed FP8$\times$FP4 differ fundamentally between Hopper and Blackwell.

\section{GC-MoE design}
\label{sec:design}

\subsection{Overview}
\label{sec:overview}

GC-MoE executes one MoE layer as one replay of a child CUDA Graph with three kernel nodes, called roles (Table~\ref{tab:roles} and Figure~\ref{fig:architecture}). Role1 (\emph{dispatch}) runs on the communication green context and performs routing accounting and the all-to-all pull into a local token ring. Role2 (\emph{compute}) runs on the compute green context and executes the routed and shared grouped GEMMs with fused SwiGLU and requantization, scattering BF16 partials directly to the source ranks. Role3 (\emph{finalize}) runs on the compute green context and combines the $k$ (plus one shared) partials of Eq.~\eqref{eq:moe} into the output. Role1 and role2 are \emph{concurrent roots} of the graph---there is no dependency edge between them (Figure~\ref{fig:architecture} draws them side by side)---so the driver schedules them simultaneously on disjoint SM sets. Role3 depends on both. Each role is an ordinary kernel with its own launch bounds, register and shared-memory footprint, and synchronization domain; no CTA budget is shared across roles. This is what lets role1 be tuned for communication latency and role2 for weight streaming and MMA throughput, and it is why a change to one path does not perturb the others. Algorithm~\ref{alg:overview} summarizes one replay.

On the SM90 path, the same structure holds with \texttt{sm\_90a} kernels and a default dispatch request of 4 SMs; when shared experts are fused, they execute as a fourth kernel node (role4) that role2 may wait on (Section~\ref{sec:se}).

All cross-rank data---inputs, routing tables, the token ring, the combine buffer, and all signals---live in one \emph{symmetric-memory} allocation created with PyTorch's distributed symmetric-memory API and rendezvoused across ranks. Every rank obtains the peer virtual addresses of every other rank's allocation and passes them to the kernels; a 592-byte by-value descriptor carries the base address and 72 relative offsets, so a kernel maps any peer's base with a signed delta. Kernels publish and consume per-expert counters, ring counters, and combine readiness through these peer addresses over NVLink.

\begin{table}[t]
\centering
\caption{Roles of the SM103 CUDA Graph (three nodes; on SM90, fused shared experts add a fourth node). \texttt{dispatch\_sms}/\texttt{compute\_sms} are the driver-aligned partition sizes. \texttt{finalize\_sms} is $\min(\texttt{compute\_sms},64)$ at up to 128 tokens per rank and \texttt{compute\_sms} above that; with fused shared experts it is additionally tightened by the combine geometry to $\max(\min(\texttt{compute\_sms},\,e+1),\,\lceil T\cdot\text{combine chunks}/8\rceil)$.}
\label{tab:roles}
\small
\begin{tabular}{llllll}
\toprule
Role & Function & Green context & Grid & Threads & Cluster \\
\midrule
role1 & token dispatch (all-to-all pull) & communication & \texttt{dispatch\_sms} & 256 or 384 & -- \\
role2 & FP8$\times$FP4 grouped GEMM + SwiGLU & compute & \texttt{compute\_sms} & 384 & (2,1,1) \\
role3 & signal reset + combine & compute & \texttt{finalize\_sms} & 384 & -- \\
\bottomrule
\end{tabular}
\end{table}

\begin{algorithm}[t]
\caption{GC-MoE layer execution (one CUDA Graph replay).}
\label{alg:overview}
\begin{algorithmic}[1]
\Require FP8 activations $x$ with packed UE8M0 scales, routing $(\mathit{idx},g)$, transformed FP4 routed weights $W_1,W_2$, symmetric buffer
\Ensure BF16 output $y$
\Statex \textit{Role1 --- dispatch kernel, communication partition (concurrent graph root)}
\State count local tokens per expert; publish per-expert counts to peers with packed atomics
\State \textbf{NVLink barrier} (tag~1); pull remote tokens, scales, and routing metadata into the token ring via TMA staging
\State release \texttt{l1\_full\_count} for each filled ring block (generation-tagged)
\Statex \textit{Role2 --- compute kernel, compute partition (concurrent graph root)}
\State scheduler warp broadcasts W1/W2/shared tasks to both CTAs of each cluster
\For{each ring block $b$ in generation order}
  \State \textbf{wait} \texttt{l1\_full\_count}[$b$] (acquire) \Comment{starts as soon as block $b$ is filled}
  \State $u\gets$ FP8$\times$FP4 GEMM $W_1$; fuse SwiGLU, routing weight, and FP8 requantization
  \State set \texttt{l2\_full\_mask} bits of finished W1 tiles; $v\gets$ GEMM $W_2$; scatter BF16 partials to source ranks
\EndFor
\Statex \textit{Role3 --- finalize kernel, compute partition (after role1 and role2)}
\State publish combine readiness to all peers; consume double-buffered receive counts; reset ring signals
\State \textbf{wait} peer readiness; gather, reduce, and write $y$ via TMA
\end{algorithmic}
\end{algorithm}

\subsection{Green-context SM partitioning}
\label{sec:partition}

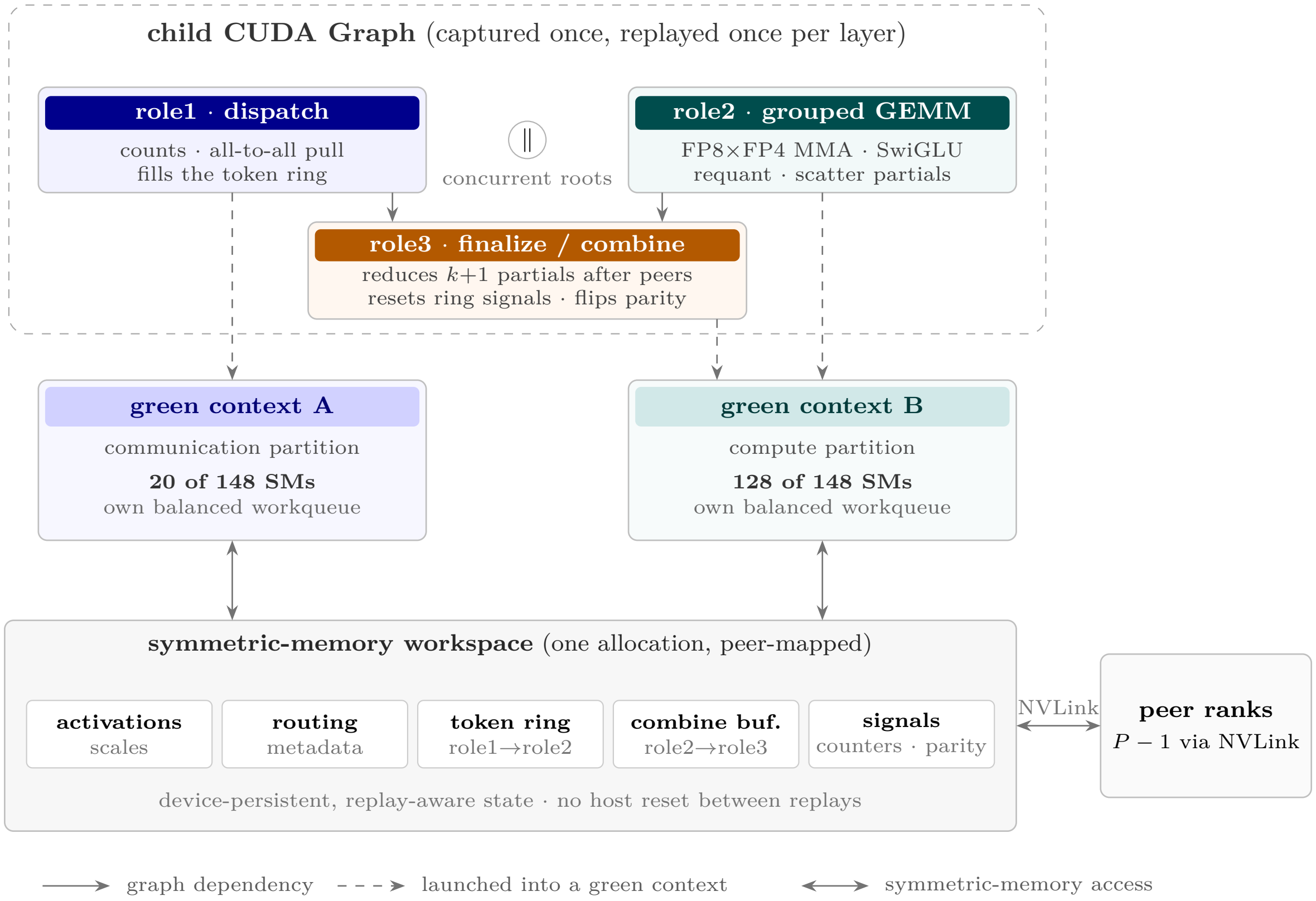
\begin{figure}[t]
\centering
\begin{tikzpicture}[font=\footnotesize, >={Stealth[length=1.8mm]},
  box/.style={rounded corners=3pt, draw=black!25, line width=0.5pt, align=center, inner sep=3pt},
  hdr/.style={rounded corners=2pt}]
\draw[black!35, dashed, rounded corners=6pt] (0.15,7.85) rectangle (12.45,11.75);
\node[font=\small, text=black!80] at (6.3,11.40)
  {\textbf{child CUDA Graph} (captured once, replayed once per layer)};
\node[box, fill=blue!5, minimum width=4.6cm, minimum height=1.25cm] (r1) at (2.8,10.15) {};
\fill[blue!55!black, hdr] (0.58,10.27) rectangle (5.02,10.67);
\node[white, font=\footnotesize\bfseries] at (2.8,10.47) {role1 $\cdot$ dispatch};
\node[font=\scriptsize, align=center, text=black!75] at (2.8,9.87)
  {counts $\cdot$ all-to-all pull\\ fills the token ring};
\node[box, fill=teal!5, minimum width=4.6cm, minimum height=1.25cm] (r2) at (9.8,10.15) {};
\fill[teal!60!black, hdr] (7.58,10.27) rectangle (12.02,10.67);
\node[white, font=\footnotesize\bfseries] at (9.8,10.47) {role2 $\cdot$ grouped GEMM};
\node[font=\scriptsize, align=center, text=black!75] at (9.8,9.87)
  {FP8$\times$FP4 MMA $\cdot$ SwiGLU\\ requant $\cdot$ scatter partials};
\node[circle, draw=black!30, fill=white, inner sep=1.3pt] at (6.3,10.15) {$\parallel$};
\node[font=\scriptsize, text=black!55] at (6.3,9.70) {concurrent roots};
\node[box, fill=orange!6, minimum width=5.2cm, minimum height=1.15cm] (r3) at (6.3,8.60) {};
\fill[orange!70!black, hdr] (3.78,8.71) rectangle (8.82,9.09);
\node[white, font=\footnotesize\bfseries] at (6.3,8.90) {role3 $\cdot$ finalize / combine};
\node[font=\scriptsize, align=center, text=black!75] at (6.3,8.40)
  {reduces $k{+}1$ partials after peers\\ resets ring signals $\cdot$ flips parity};
\draw[->, black!55, line width=0.5pt] (4.7,9.525) -- (4.7,9.175);
\draw[->, black!55, line width=0.5pt] (7.9,9.525) -- (7.9,9.175);
\node[box, fill=blue!3, minimum width=4.6cm, minimum height=1.9cm] (ga) at (2.8,6.35) {};
\fill[blue!18, hdr] (0.58,6.75) rectangle (5.02,7.22);
\node[font=\footnotesize\bfseries, text=blue!45!black] at (2.8,6.98) {green context A};
\node[font=\scriptsize, text=black!70] at (2.8,6.48) {communication partition};
\node[font=\scriptsize\bfseries, text=black!80] at (2.8,6.10) {20 of 148 SMs};
\node[font=\scriptsize, text=black!60] at (2.8,5.78) {own balanced workqueue};
\node[box, fill=teal!3, minimum width=4.6cm, minimum height=1.9cm] (gb) at (9.8,6.35) {};
\fill[teal!18, hdr] (7.58,6.75) rectangle (12.02,7.22);
\node[font=\footnotesize\bfseries, text=teal!45!black] at (9.8,6.98) {green context B};
\node[font=\scriptsize, text=black!70] at (9.8,6.48) {compute partition};
\node[font=\scriptsize\bfseries, text=black!80] at (9.8,6.10) {128 of 148 SMs};
\node[font=\scriptsize, text=black!60] at (9.8,5.78) {own balanced workqueue};
\draw[->, dashed, black!55, line width=0.5pt] (2.8,9.525) -- (2.8,7.30);
\draw[->, dashed, black!55, line width=0.5pt] (9.8,9.525) -- (9.8,7.30);
\draw[->, dashed, black!55, line width=0.5pt] (8.55,8.025) -- (8.55,7.30);
\node[box, fill=gray!6, minimum width=12.0cm, minimum height=2.5cm] (sym) at (6.1,3.2) {};
\node[font=\footnotesize, text=black!85] at (6.1,4.16)
  {\textbf{symmetric-memory workspace} (one allocation, peer-mapped)};
\foreach \a/\b/\c in {1.46/activations/scales, 3.78/routing/metadata,
                      6.10/token ring/role1$\to$role2, 8.42/combine buf./role2$\to$role3,
                      10.74/signals/counters $\cdot$ parity}{
  \node[draw=black!20, fill=white, rounded corners=2pt, minimum width=2.2cm,
        minimum height=0.8cm, align=center, inner sep=1pt] at (\a,3.10)
        {\scriptsize\textbf{\b}\\[-1pt]{\scriptsize\color{black!55}\c}};
}
\node[font=\scriptsize, text=black!55] at (6.1,2.30)
  {device-persistent, replay-aware state $\cdot$ no host reset between replays};
\draw[<->, black!55, line width=0.5pt] (2.8,5.40) -- (2.8,4.45);
\draw[<->, black!55, line width=0.5pt] (9.8,5.40) -- (9.8,4.45);
\node[box, fill=gray!4, minimum width=2.5cm, minimum height=1.7cm] (peer) at (14.35,3.2)
  {\textbf{peer ranks}\\[1pt]{\scriptsize $P-1$ via NVLink}};
\draw[<->, black!55, line width=0.5pt] (12.1,3.2) -- (13.1,3.2);
\node[font=\scriptsize, text=black!55] at (12.6,3.42) {NVLink};
\draw[->, black!55, line width=0.5pt] (0.55,1.30) -- (1.35,1.30);
\node[anchor=west, font=\scriptsize, text=black!65] at (1.42,1.30) {graph dependency};
\draw[->, dashed, black!55, line width=0.5pt] (4.05,1.30) -- (4.85,1.30);
\node[anchor=west, font=\scriptsize, text=black!65] at (4.92,1.30) {launched into a green context};
\draw[<->, black!55, line width=0.5pt] (9.55,1.30) -- (10.35,1.30);
\node[anchor=west, font=\scriptsize, text=black!65] at (10.42,1.30) {symmetric-memory access};
\end{tikzpicture}
\caption{Architecture of one GC-MoE layer replay. The child CUDA Graph holds three nodes: role1 (dispatch) and role2 (grouped GEMM) have no dependency edge and are concurrent roots, whereas role3 (finalize/combine) depends on both (solid arrows). Dashed arrows map each role to the green context that launches it---role1 to the 20-SM communication context, role2 and role3 to the 128-SM compute context---each with its own balanced workqueue so that the driver co-schedules them on disjoint SM sets. All cross-rank state lives in one symmetric-memory workspace whose boxes name the buffers exchanged over NVLink (double arrows): activations and scales, routing metadata, the role1$\rightarrow$role2 token ring, the role2$\rightarrow$role3 combine buffer, and the counters and parity signals. Every synchronization object is device-persistent and replay-aware, so a replay needs no host-side reset.}
\label{fig:architecture}
\end{figure}

The runtime creates a \emph{green-context pair} once per CUDA context and caches it (Figure~\ref{fig:architecture}). Creation proceeds as follows. (1)~Query the device's SM resource, which reports the SM count, the minimum partition size, and the co-scheduling alignment. (2)~Compute the compute-target SM count as $\lfloor(\text{total}-\text{requested})/\text{alignment}\rfloor\cdot\text{alignment}$, where \emph{requested} is the application's dispatch request (\texttt{GC\_MOE\_DISPATCH\_SMS}, default 16), and split the SM resource with \texttt{cuDevSmResourceSplitByCount}; the driver returns the compute resource and the complementary communication resource. The partition is thus driver-aligned, not application-chosen: on B300 (148 SMs) the default request yields a \textbf{20/128} communication/compute split. (3)~If balanced workqueues are enabled (default), obtain a workqueue-config resource from the primary context with \texttt{sharingScope=CU\_WORKQUEUE\_SCOPE\_GREEN\_CTX\_BALANCED} and \texttt{wqConcurrencyLimit=1}, and generate the context descriptor from both resources; after creation, read the resource back and verify that the balanced limit took effect. Balanced workqueues matter in practice: without them, two contexts can be placed on one hardware workqueue, which silently serializes the very kernels the design intends to overlap. (4)~Create a non-blocking green stream per context and a retirement event; teardown records the event, synchronizes, and only then destroys the context. The pair is cached thread-locally per CUDA context. The partition is validated at runtime: at least 32 compute SMs, an even compute count, at least two dispatch SMs, and communication plus compute equal to the device SM count.

Two consequences deserve emphasis. First, the partition is static for the lifetime of a graph, which makes replay deterministic: no kernel is ever migrated, and the dispatch partition's SM count is a compile-time property of the launched kernels. Second, the cost of the partition is small in the regime that matters: the split assigns 20 of 148 SMs (13.5\%) to dispatch and 128 to compute, but at decode batch sizes the grouped GEMM is bound by weight streaming and small-$M$ issue latency rather than tensor-core throughput, so 128 SMs suffice; the measured effective throughput confirms that the smaller compute partition still beats the baseline's shared-SM GEMM (Section~\ref{sec:results}). The partition also shrinks the synchronization domain: role1's grid syncs, including those around the tag-1 barrier, span only the 20 dispatch SMs; role3's grid sync spans its finalize grid (64 SMs at up to 128 tokens per rank); and the compute partition's task scheduler is local to its own SMs, whereas the baseline's prologue and tail grid syncs span all 148.

\subsection{Overlapped dispatch and expert computation}
\label{sec:overlap}

\paragraph{Token ring.} Dispatch does not allocate per-expert buffers. All pulled tokens land in one globally reused \emph{token ring} of ring blocks of $B_M$ tokens; each ring block carries four persistent signals: \texttt{l1\_full\_count} (published by dispatch when the block is filled), \texttt{l1\_empty\_count} (released by compute when the block's activations are consumed), \texttt{l2\_full\_mask} (a 64-bit bitmap of finished W1 output tiles), and \texttt{l2\_empty\_count} (released when the block's L2 activations are consumed). Ring capacity is derived from a maximum-in-flight model: with per-rank capacity $T_{\max}$, the maximum routed assignments are $T_{\max}\cdot P\cdot\min(k,e)$; the pool of logical blocks is $\lceil \text{assignments}/B_M\rceil + e$, and the number of simultaneously live blocks is recomputed from the actual compute-partition SM count and L1/L2 cluster waves. The capacity check uses the real partition size, so a graph never starts with an undersized ring. Because the ring is reused across generations of blocks, each logical pool block maps to a ring block plus a generation number; dispatch must observe that the ring block's consumers have released the previous generation ($\texttt{l1\_empty\_count}$) before overwriting it. Generation tags make block reuse safe across generations within a replay; between replays, role3 resets the ring signals in-kernel, overlapped with the combine half (Section~\ref{sec:combine}), so no separate reset pass is needed.

\paragraph{Producer--consumer handshake.} The compute kernel's TMA-producer warp waits with acquire semantics on $\texttt{l1\_full\_count} = B_M\cdot(\text{generation}+1)$ for the target ring block before publishing the block's W1 tasks. The handshake granularity is thus one ring block: the GEMM starts on the first filled block, not at the end of dispatch. Within a block, the L2 stage does not need a separate barrier either: each W2 K128 tile consumes two W1 N128 output tiles, and the W1 epilogue sets the corresponding bits in the 64-bit \texttt{l2\_full\_mask}; the L2 producer waits for exactly its two bits. The bitmap keeps one bit per W1 N128 tile, and each W2 K128 tile waits on exactly its two bits; this is also why $2I/128\leq 64$ is a hard design constraint (Appendix~\ref{app:constraints}).

\paragraph{Scheduler.} Each cluster's leader CTA runs a scheduler warp. Tasks are phases over blocks: Linear1, Linear2, SharedLinear1, SharedLinear2. The scheduler pre-publishes a small number of L1 waves (computed from the same ring model) before the first L2 wave, then maintains a one-in-one-out discipline: every successful L2 claim publishes one more L1 task, and an L2 task is published only after all L1 tasks of its pool block have been published. Shared-expert tasks are injected before routed L2 work when small (interleaved with routed tasks) and queued after routed work when large. Task records are 32-byte descriptors broadcast to the two CTAs of a cluster through distributed shared memory with transaction-count mbarriers and double buffering.

\paragraph{Dispatch data path.} Role1 proceeds in six steps. (1)~File-level token counting: each warp processes $32/k$ tokens per iteration and accumulates per-expert counts in shared memory. (2)~A packed atomic add of $(1\ll32)\,|\,\mathit{count}$ to the local per-expert send counter; the high word counts the number of (rank, dispatch-SM) contributors for arrival validation, the low word the token count. (3)~A second scan writes each token's routing entry to the \emph{target} rank's source-index region (peer write). (4)~After a grid-wide sync, block~0 publishes the per-expert counts and prefix sums to every peer's receive counters. (5)~A two-phase NVLink barrier (tag~1) ensures all peers' counts are visible before pulls begin; the barrier is a persistent, phase-encoded counter that survives graph replays, and it carries a configurable timeout (default 3600\,s) after which the device traps instead of hanging. (6)~Each warp walks the expert-concatenated, unpadded token sequence with a global cursor: lane-held per-rank count registers plus a warp reduction locate each token's $(\mathit{src\_rank},\mathit{src\_slot})$, and warp-private double-buffered shared-memory staging performs chunked bulk copies of the FP8 activation from the peer and the corresponding packed-scale rows, permuted to the MN-major row order the MMA path expects. Routing weight and $(\mathit{src\_rank},\mathit{src\_token},\mathit{slot})$ metadata are written alongside; the block's full count is released when the last token lands.

\paragraph{No trailing barrier.} Early designs ended each invocation with a second NVLink barrier---DeepGEMM's Mega MoE still ends its kernel this way, after workspace cleaning---so that the next replay could safely reuse the receive counters and workspace. GC-MoE instead makes the per-expert receive counters \emph{double-buffered by parity}: a replay consumes generation $N\bmod 2$, and role3 flips the parity after consuming the counts and completing its grid sync, so the next publish targets the other generation. Ordering is guaranteed by the tag-1 barrier and a release/acquire chain on combine readiness, not by a cross-rank barrier. This removes a full NVLink round trip from the inter-replay critical path.

\paragraph{Direct partial scatter.} The W2 epilogue does not write intermediate results to a local buffer for a later combine kernel. Each lane issues 16-byte stores of BF16 fragments directly to the \emph{source rank}'s combine buffer at the slot $(\mathit{topk\_slot},\mathit{src\_token})$; buffer recycling is governed by the \texttt{l1\_empty}/\texttt{l2\_empty} handshake. The combine kernel therefore only has to reduce, not gather from a local staging area.

\subsection{\texorpdfstring{FP8$\times$FP4}{FP8xFP4} grouped GEMM}
\label{sec:gemm}

\paragraph{Blackwell path (SM103).} On B300 the operator uses the 2-CTA block-scaled MMA \texttt{tcgen05.}\allowbreak\texttt{mma.}\allowbreak\texttt{cta\_group::2.}\allowbreak\texttt{kind::mxf8f6f4.}\allowbreak\texttt{block\_scale}: two CTAs of a cluster jointly issue one MMA whose FP32 accumulator lives in TMEM and is read back with \texttt{tcgen05.ld} after the MMA-completion mbarrier. As in the baseline, the operands are swapped: the instruction's A operand is the weight tile and B the activation tile, so the MMA's N dimension is the token dimension. The instruction N is set to the tile's valid token count rounded up to 16 rather than the padded block size, so partially filled blocks cost proportionally less. Because A carries the varying dtype, the same instruction sequence serves FP4 routed weights (E2M1, format~5) and FP8 shared weights (E4M3, format~0) with a runtime-selected format id; the CuTe DSL's public block-scaled MMA requires matching A/B dtypes, so the instruction is emitted as an inline-PTX user op. The scale factors are moved with the dedicated UTCCP path: per K128 stage, the activation scales ($B_M$-aligned rows, four packed K32 groups per 16 bytes) are copied to TMEM with \texttt{tcgen05.cp.cta\_group::2.32x128b.warpx4}, 512 bytes per stage; weight scales take the same path. Each K128 stage is then issued as four K32 MMAs whose scale-factor byte selector picks the corresponding K32 group; for the $(1,128,128)$ shared recipe, weight scales are staged once per four K128 blocks and the selector is driven by the K128 block index modulo 4. The Blackwell-specific contribution is therefore not the instruction pattern---operand swap, valid-M N sizing, and the UTCCP scale path follow the baseline---but the scheduling around it: K128 tile-prepacked weights, the shared-expert task stream, and the partition-level overlap.

Tile geometry is fixed at $B_N=B_K=128$, with each CTA of the pair responsible for half of the $B_M$ rows. The TMEM allocation is $2B_M$ accumulator columns plus $\lceil B_M/128\rceil$-derived scale columns, rounded up to a power of two and capped at 512 columns. The 64-bit W1-completion bitmap caps the W1 output-tile count at 64, i.e., $2I/128\leq 64$; the even-$H/128$ requirement follows from pairing N128 tiles per cluster.

\paragraph{Weight prepacking.} The weight transform rewrites each expert's quantized FP4 matrix from row-major $[E,\mathit{rows},K/2]$ into K128-tile-contiguous $[E,k_b,\mathit{rows},64]$: within a tile, all rows' 64-byte segments are contiguous. A TMA descriptor encodes the 4-D box $(128,128,1,1)$ over $(K,\mathit{rows},k_b,\mathit{experts})$, so one TMA operation bursts a single 8\,KB contiguous global range (16\,KB for FP8 shared weights) instead of 128 segments spaced a full weight row apart; the FP4 payload is unpacked from 16U4-aligned storage into 8-bit shared memory (SMEM) by the tensor map. Gate and up rows are interleaved in 8-row groups so the SwiGLU epilogue reads pairs from one tile, and scale rows are permuted to the UTCCP consumption order; both follow the baseline's weight transform. The K128 tile packing is the addition: it turns each weight tile load into one contiguous burst instead of 128 segments spaced a full weight row apart. The transform runs asynchronously on the caller's stream once per weight set and is reused across token counts, capacities, and graph captures. Routed FP4 weight loads keep the baseline's \texttt{EVICT\_FIRST} cache hint: decode streams the weights once per layer, and polluting L2 with them hurts the activation/scales working set.

\paragraph{W1 epilogue.} Accumulators are read from TMEM in FP32, rounded to BF16, clamped, activated with the sigmoid term of SwiGLU (optionally using a polynomial exp and reciprocal for fast math), multiplied by the up projection and by the token's routing weight (applied in the W1 stage so the L2 GEMM needs no per-token scaling), reduced to a warp-group amax, converted to a UE8M0 scale and an E4M3 payload, and stored through SMEM with a transposing stmatrix to the L2 activation ring via TMA; the requantized scale rows are written directly in the permuted order that the next stage's UTCCP expects. At $B_M=240$ the epilogue double-buffers registers so the TMEM read of the next store overlaps the arithmetic of the current one.

\paragraph{Block-size heuristic.} The host picks $B_M$ from the expected tokens per expert $\hat{m}=T\cdot P\cdot k/E$ and its standard deviation, covering $\hat{m}+\sqrt{\hat{m}}$ with the number of blocks at the largest candidate; the candidate set is $\{16,32,64,96,128,192,240\}$, small batches pin $B_M=16$--$32$, and at least two blocks are used when the per-rank expert count is small. The chosen value balances MMA granularity against the number of live ring blocks.

\paragraph{Hopper path (SM90).} The SM90 operator is an independent implementation. Activations are E4M3FN with contiguous FP32 K128 scales; routed weights are MXFP4 (E2M1 with UE8M0 K32 scales); shared weights are FP8. Two mechanisms replace the missing hardware blockscale path. First, the host transform repacks MXFP4 weights into WGMMA-friendly tiles using a \emph{window-based} rescaling: for each expert it computes a window base from the scale extremes (at most 11 exponent steps) and shifts FP4 magnitudes that fall outside the window down by exact powers of two using a lookup table over the 4-bit code space, recording the per-expert compensating factor; the tile bytes are simultaneously permuted to the lane semantics the register decoder expects. Second, the kernel decodes FP4 to E4M3 \emph{in registers} with two LDS.128 loads plus PRMT/LOP3 bit manipulation, then issues register-A WGMMA \texttt{m64nNk32} with E4M3 operands; block scales are applied as per-K32 fused multiply-adds outside the MMA, and a global $\times 64$ convention on stored scales moves E4M3 values into the WGMMA e4m3 dynamic range. Tile sizes and warp specialization are chosen by a token-bound heuristic ($B_M\in\{8,16,32,64\}$, 384-thread pipelined or 256-thread non-pipelined variant), and dispatch uses epoch/counter headers with a peer table; the default dispatch request on Hopper is 4 SMs. Table~\ref{tab:paths} summarizes the two paths.

\begin{table}[t]
\centering
\caption{SM103 (Blackwell) and SM90 (Hopper) implementation paths.}
\label{tab:paths}
\small
\begin{tabular}{@{}L{0.26\linewidth}L{0.34\linewidth}L{0.34\linewidth}@{}}
\toprule
 & Blackwell (SM103, B300) & Hopper (SM90, H20) \\
\midrule
Compile target & \texttt{sm\_103f} & \texttt{sm\_90a} \\
MMA & tcgen05 \texttt{mxf8f6f4.}\allowbreak\texttt{block\_scale}, 2-CTA & register-A WGMMA \texttt{m64nNk32}, E4M3 \\
Weight format & packed FP4 with UE8M0 K32 scales & MXFP4, window-repacked with per-expert factor \\
Activation format & E4M3 with packed UE8M0 K32 scales & E4M3FN with FP32 K128 scales \\
FP4 handling & native E2M1 operand (format 5) & in-register decode FP4$\rightarrow$E4M3 \\
Scale application & UTCCP to TMEM; hardware block scale & per-K32 FMA outside MMA; $\times 64$ headroom \\
Weight layout & K128 tile-contiguous prepack & windowed tile repack, lane-permuted \\
Dispatch request & \texttt{GC\_MOE\_}\allowbreak\texttt{DISPATCH\_SMS}=16 (20 aligned) & \texttt{GC\_MOE\_SM90\_}\allowbreak\texttt{DISPATCH\_SMS}=4 \\
Graph nodes & 3 & 3 (4 with fused shared experts) \\
\bottomrule
\end{tabular}
\end{table}

\subsection{Fused shared experts}
\label{sec:se}

Shared experts are optional and fused into the same task stream rather than run as separate layers. The shared-expert weights are FP8 in K128 packed tiles. Three details make the fusion cheap. First, the shared W1 GEMM reads the \emph{original} activation buffer by aliasing it (zero copy): shared experts see every local token, so no dispatch is needed. Second, shared activation scales must be presented in the same MN-major permuted layout that the routed path's UTCCP consumes, although they are always K32 regardless of the shared weight recipe. The caller stages them once per token count into \texttt{shared\_l1\_acts\_sf}: for token $t$ with $u=t\bmod B_M$, the destination row is
\begin{equation}
\mathrm{row}(t)=\left\lfloor t/B_M\right\rfloor\cdot\mathrm{align}(B_M,128)+\left\lfloor u/128\right\rfloor\cdot128+(u\bmod 32)\cdot4+\left\lfloor (u\bmod 128)/32\right\rfloor .
\label{eq:staging}
\end{equation}
The row order of Eq.~\eqref{eq:staging} is exactly the one the kernel itself writes for routed scales, so shared and routed experts share one TMA scale map and one UTCCP path; the buffer holds $\lceil T_{\max}/8\rceil\cdot128$ rows, an upper bound valid for every candidate $B_M$. The staging can be fused into an upstream producer (e.g., the quantizer). Third, shared tasks enter the unified scheduler as SharedLinear1/SharedLinear2 phases and are placed before or after routed work depending on size, and the shared output occupies lane $k$ of the combine buffer, so the finalize kernel reduces it with no special case. On SM90, shared experts run as a fourth graph node (role4) that role2 optionally waits on.

\subsection{Three-node CUDA Graph combine}
\label{sec:combine}

Role3 is a single kernel with two halves that run concurrently. The first four warps perform the \emph{bookkeeping} half: publish combine readiness to every peer with a release system atomic (required even for ranks with no routed tokens), consume this generation's per-expert receive counts with an arrival check against $\texttt{dispatch\_sms}\cdot P$, flip \texttt{recv\_parity}, and reset the ring signals, counters, and the W1 bitmap for the next replay, accumulating optional per-expert receive statistics along the way. Because role3 depends on both graph roots, this half starts after role2 completes, but it does not add a separate serial phase: it runs alongside the combine half, and its resets proceed while the combine warps wait for peer readiness and reduce. The remaining warps perform the \emph{combine} half: wait for each peer's readiness with a bounded timeout, then, for each token chunk assigned to the partition, prefetch and accumulate the valid top-$k$ (and shared) slots from the possibly remote combine buffer with double-buffered bulk copies, accumulate BF16 values into FP32 registers, pack pairs back to BF16, and write $y$ through SMEM and a bulk TMA store. The chunk geometry (1, 2, or 4 chunks per token) is chosen from the reusable SMEM budget with 16-byte per-lane alignment.

The three-node formulation is deliberate. Role1 and role2 have no dependency edge, which is what makes the driver run them concurrently. Role3 depends on both and thereby absorbs the global completion condition---all W2 partials scattered---through release/acquire readiness signals rather than a second cross-rank barrier. And because every synchronization object is device-persistent and replay-aware (barrier phases, ring generations, receive parity, combine readiness), a graph replay needs no host-side reset. The child graph is attached to the caller's capture with capture-dependency semantics, or launched directly as a graph outside capture; it is cached by cubin hash, parameter signature, and resource state, so repeated layers and repeated token counts replay the same graph.

\section{Implementation}
\label{sec:impl}

GC-MoE is written in Python against NVIDIA's CuTe DSL (CUTLASS DSL 4.8.1) and JIT-compiled per configuration to PTX, which the system assembler (\texttt{ptxas} 13.2) turns into cubin. The operator is a Python package; installing it builds a wheel but compiles no kernels---compilation happens on first use of a configuration.

\paragraph{Compiler identity and caching.} A compiler-identity record binds the ptxas path, size, version, and hash; the DSL frontend hash; a content hash of the operator's own Python source tree; native-library identity; and the effective options. Options are locked to \texttt{-{}-gpu-arch <target> -{}-opt-level 3} with a whitelist for ptxas flags (we use \texttt{-O3}, register-usage level 10, and fused multiply-add); relocatable device code and user overrides are rejected. Two caches sit above the assembler: a binary cache keyed by compiler identity plus kernel attributes, resources, and tensor metadata (never CUDA addresses), and a capture-key cache that additionally keys tensor-map encodings and loaded kernels by context, green partition, and all pointer/shape/timeout arguments. Cubins are loaded through the driver's library API and reused by hash.

\paragraph{By-value kernel parameters.} CUDA Graph nodes hold kernel parameters by value, which is essential here: the compute kernel needs 18 tensor maps (TMA descriptors for activations, scales, and weights) and the peer address table. CuTe emits both as pointer formals. A post-generation PTX rewrite converts the peer descriptor into a 592-byte by-value \texttt{.b8} formal (base, 72 int64 offsets, rank index) and each tensor map into a 128-byte aligned by-value formal, rewriting every param-space address in \texttt{prefetch.tensormap} and \texttt{cp.async.bulk.tensor} instructions through \texttt{cvta.param.u64}; the rewrite fails closed if the generated PTX shape changes, and the parameter order is normalized for the graph ABI. The SM90 path shares the JIT utilities and validates its ABI (14 pointers plus 3 scalars, and the presence of \texttt{wgmma.mma\_async} and \texttt{cp.async.bulk}).

\paragraph{Distributed interface.} Initialization requires one process per GPU, peer-accessible GPUs, and a torch distributed group. The buffer API allocates and rendezvouses the symmetric memory; the weight-transform API returns transformed weight pairs once per weight set and may be called outside graph capture, asynchronously on the caller's stream. The operator call writes the output in place. Listing~\ref{lst:usage} shows the routed-expert invocation; fused shared experts extend it with shared weight pairs and the staged scale buffer.

\begin{lstlisting}[caption={Minimal GC-MoE invocation (Blackwell path, routed experts).},label={lst:usage}]
buffer = gc_moe.get_symm_buffer_for_gc_moe(
    group, num_experts, capacity, topk, hidden, intermediate,
    mma_type='fp8xfp4')
w1, w2 = gc_moe.transform_weights_for_gc_moe(l1_weights, l2_weights)

buffer.x[:tokens].copy_(x_fp8)
buffer.x_sf[:tokens].copy_(x_sf)
buffer.topk_idx[:tokens].copy_(topk_idx)
buffer.topk_weights[:tokens].copy_(topk_weights)
y = torch.empty((tokens, hidden), dtype=torch.bfloat16,
                device=x_fp8.device)
gc_moe.fp8_fp4_gc_moe(y, w1, w2, buffer, recipe=(1, 1, 32),
                      activation='swiglu', activation_alpha=1.0,
                      activation_clamp=10.0)
\end{lstlisting}

\paragraph{Size and constraints.} The device code is 1935 lines for the Blackwell (SM103) roles and 1305 lines for the Hopper (SM90) roles, plus shared scheduler, barrier, TMA, and PTX helpers. The operator's shape constraints and their technical reasons are collected in Appendix~\ref{app:constraints}; the most important are $I\leq 4096$ with $I$ a multiple of 128 (64-bit W1 bitmap and K128 tiles), $H$ a multiple of 256 (2-CTA N-tile pairing), $k+\mathbf{1}[\text{shared}]\leq 32$ (one warp lane per combine slot), and at most 2048 experts globally (static signal sizing).

\section{Evaluation}
\label{sec:eval}

\subsection{Setup}
\label{sec:setup}

\paragraph{Environment.} Table~\ref{tab:env} lists hardware and software. We measure two generations: eight NVIDIA H20 GPUs (Hopper, \texttt{sm\_90a}) and eight NVIDIA B300 GPUs (Blackwell, compute capability 10.3, \texttt{sm\_103f}). The baseline is DeepGEMM's Mega MoE~\citep{deepgemm}, measured with the \texttt{deep-gemm} package at 2.8.2+\texttt{fcad603}. On B300 we use the public \texttt{fp8\_fp4\_mega\_moe}; when shared experts are enabled, the baseline is invoked in its fused shared-expert mode (\texttt{num\_shared\_experts=1}), so that comparison is feature-matched. The public release ships only the SM100 kernel; for H20 we built an SM90 reference port of the same design, \texttt{sm90\_fp8\_fp4\_mega\_moe} (one persistent fused kernel with warp-specialized dispatch, non-epilogue, and epilogue threads around the token ring, with NVLink barriers). It has no fused shared-expert path, so the H20 shared-expert baseline runs the routed Mega kernel plus a separate shared-expert kernel.

\paragraph{Workloads.} Table~\ref{tab:models} lists the two production-scale MoE configurations. GLM-5.2 is evaluated on B300 and DeepSeek-V4.1-Flash on both hardware generations. Both use SwiGLU with $\alpha=1.0$ and clamp 10.0, the $(1,1,32)$ routed recipe, fast-math sigmoid, and a per-rank token capacity of 256.

\paragraph{Protocol.} Both operators are invoked under CUDA Graph replay with \emph{identical} inputs, capacities, weight transformations, and measurement settings. Each measurement window replays the graph 100 times; the per-invocation latency is the CUDA-event elapsed time divided by 100. We run 10 warmup windows and 30 measured windows, flushing L2 and synchronizing between windows. Each rank reports the window-median latency; the latency of a point is the \emph{maximum} across the eight ranks (the slowest rank). A decode speedup is the per-point ratio $\text{latency}_{\mathrm{baseline}}/\text{latency}_{\mathrm{GC}}$; the reported mean is the arithmetic mean of the per-point ratios (not the ratio of means). All runs are gated on idle GPUs (0\% utilization, no compute processes, at least 51\,200\,MiB free, observed over at least three samples spanning ten seconds). We scan $T\in\{1,2,4,8,16,32,64,128,256\}$ tokens per rank, i.e., 8--2048 tokens in total across the eight ranks.

\begin{table}[t]
\centering
\caption{Evaluation environment.}
\label{tab:env}
\small
\begin{tabular}{L{0.17\linewidth}L{0.77\linewidth}}
\toprule
GPUs & 8$\times$ NVIDIA H20 (Hopper, \texttt{sm\_90a}); 8$\times$ NVIDIA B300 (Blackwell, CC 10.3) \\
CUDA / assembler & CUDA 13.2, ptxas 13.2.78 \\
PyTorch & 2.11.0 (cu130) \\
Kernel DSL & NVIDIA CUTLASS DSL (CuTe) 4.8.1 \\
Baseline & DeepGEMM 2.8.2+\texttt{fcad603} (\texttt{fp8\_fp4\_mega\_moe} on B300, fused shared experts when enabled; our SM90 reference port \texttt{sm90\_fp8\_fp4\_mega\_moe} on H20) \\
Proposed & GC-MoE 0.3.1 (commit \texttt{9bdb81f}) \\
Execution & one process per GPU; PyTorch symmetric memory over NVLink; CUDA Graph replay \\
\bottomrule
\end{tabular}
\end{table}

\begin{table}[t]
\centering
\caption{Model configurations. Routed recipe $(1,1,32)$ for both models; shared-weight K group and shared recipe as shown.}
\label{tab:models}
\small
\begin{tabular}{lrrrrrr}
\toprule
Model & $H$ & $I$ & $E$ & top-$k$ & shared experts & shared K group \\
\midrule
GLM-5.2 & 6144 & 2048 & 256 & 8 & 1 & 128 \\
DeepSeek-V4.1-Flash & 5120 & 2304 & 384 & 6 & 1 & 32 \\
\bottomrule
\end{tabular}
\end{table}

\subsection{Decode latency and speedup}
\label{sec:results}

\begin{figure}[t]
\centering
\includegraphics[width=0.92\linewidth]{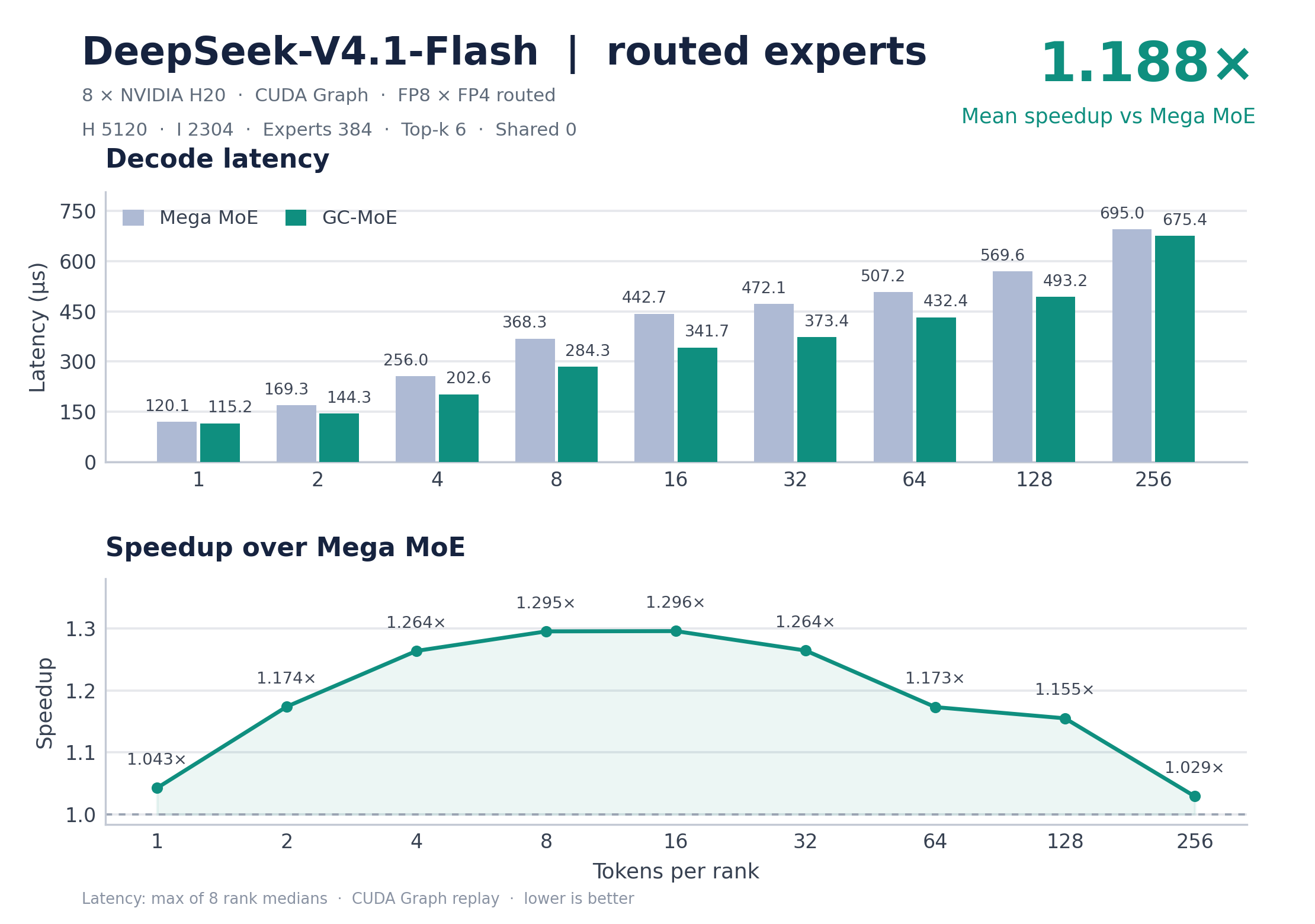}
\caption{DeepSeek-V4.1-Flash routed-expert decode on 8$\times$H20: latency and speedup of GC-MoE versus Mega MoE.}
\label{fig:h20}
\end{figure}

\paragraph{Hopper (8$\times$H20).} Table~\ref{tab:h20} and Figure~\ref{fig:h20} report DeepSeek-V4.1-Flash decode. GC-MoE is faster at all nine points, with a mean speedup of 1.188$\times$ and a peak of 1.296$\times$ at 16 tokens per rank. The shape of the curve follows the motivation: the speedup rises from 1.04$\times$ at 1 token per rank to the 1.26--1.30$\times$ plateau at 4--32 tokens per rank, where there is enough dispatch traffic for the dedicated partition to matter yet the grouped GEMM is still issue-bound at small $M$, so the baseline's shared-SM carve-out and the $\max(T_{\mathrm{combine}},\,t_{\mathrm{clean}}+t_{\mathrm{tag2}})$ tail of Eq.~\eqref{eq:mega} weigh on a short critical path; it then narrows to 1.03$\times$ at 256 tokens per rank, where weight streaming dominates and both operators approach their streaming limit, so the saving shrinks to the roughly fixed shared-SM and tail costs.

\begin{table}[t]
\centering
\caption{Decode latency and speedup, DeepSeek-V4.1-Flash on 8$\times$H20 (routed experts). Slowest-rank median per-invocation latency under CUDA Graph replay.}
\label{tab:h20}
\small
\begin{tabular}{rrrrr}
\toprule
Tokens/rank & Total tokens & Mega ($\mu$s) & GC-MoE ($\mu$s) & Speedup \\
\midrule
1 & 8 & 120.061 & 115.161 & 1.0425 \\
2 & 16 & 169.340 & 144.259 & 1.1739 \\
4 & 32 & 256.018 & 202.589 & 1.2637 \\
8 & 64 & 368.254 & 284.310 & 1.2953 \\
16 & 128 & 442.722 & 341.689 & \best{1.2957} \\
32 & 256 & 472.090 & 373.393 & 1.2643 \\
64 & 512 & 507.207 & 432.383 & 1.1731 \\
128 & 1024 & 569.604 & 493.194 & 1.1549 \\
256 & 2048 & 695.010 & 675.370 & 1.0291 \\
\midrule
\multicolumn{3}{r}{Mean speedup (9 points)} & -- & \best{1.1881} \\
\bottomrule
\end{tabular}
\end{table}

\begin{figure}[t]
\centering
\includegraphics[width=0.92\linewidth]{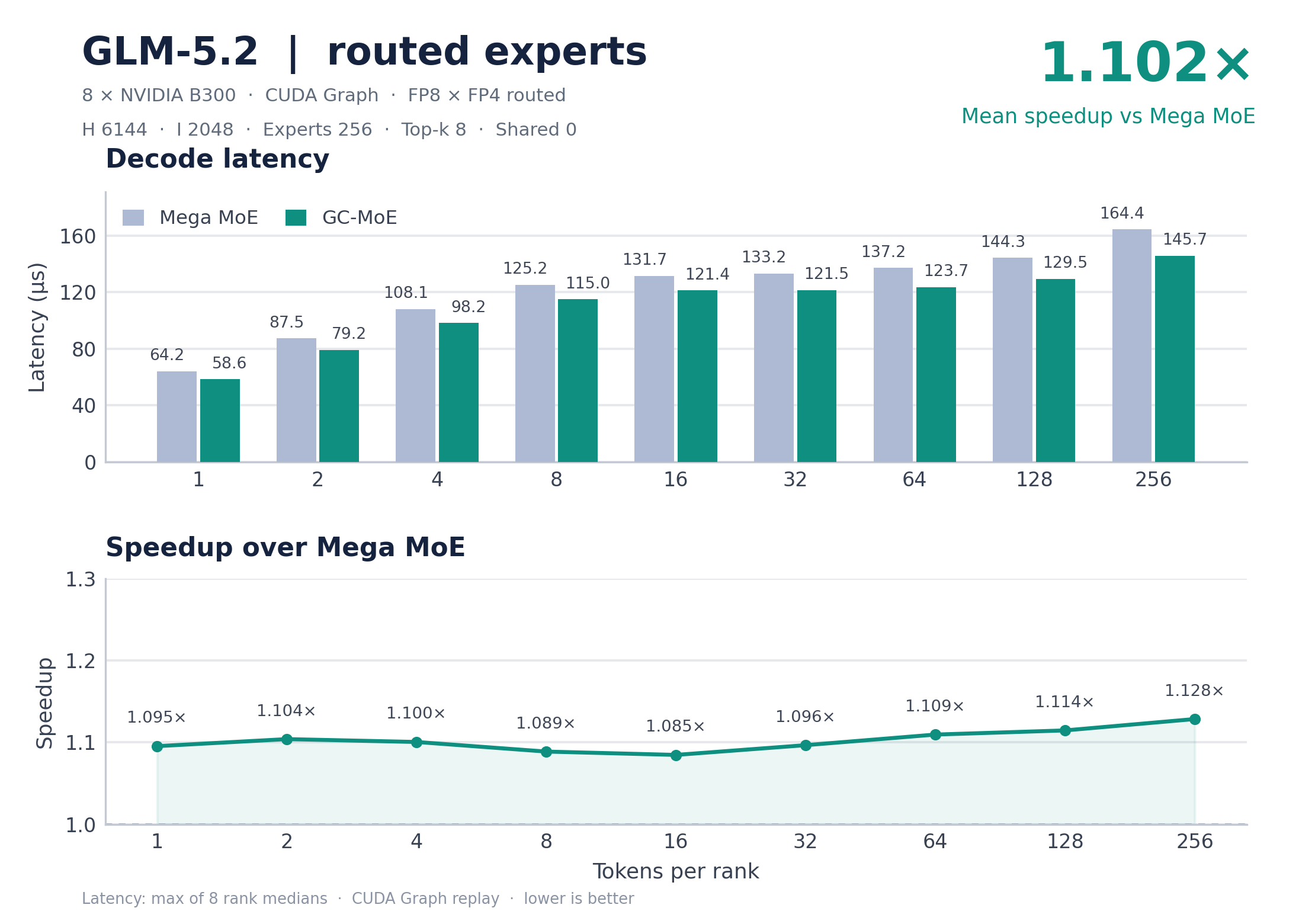}
\caption{GLM-5.2 routed-expert decode on 8$\times$B300: latency and speedup.}
\label{fig:b300glm}
\end{figure}

\begin{figure}[t]
\centering
\includegraphics[width=0.92\linewidth]{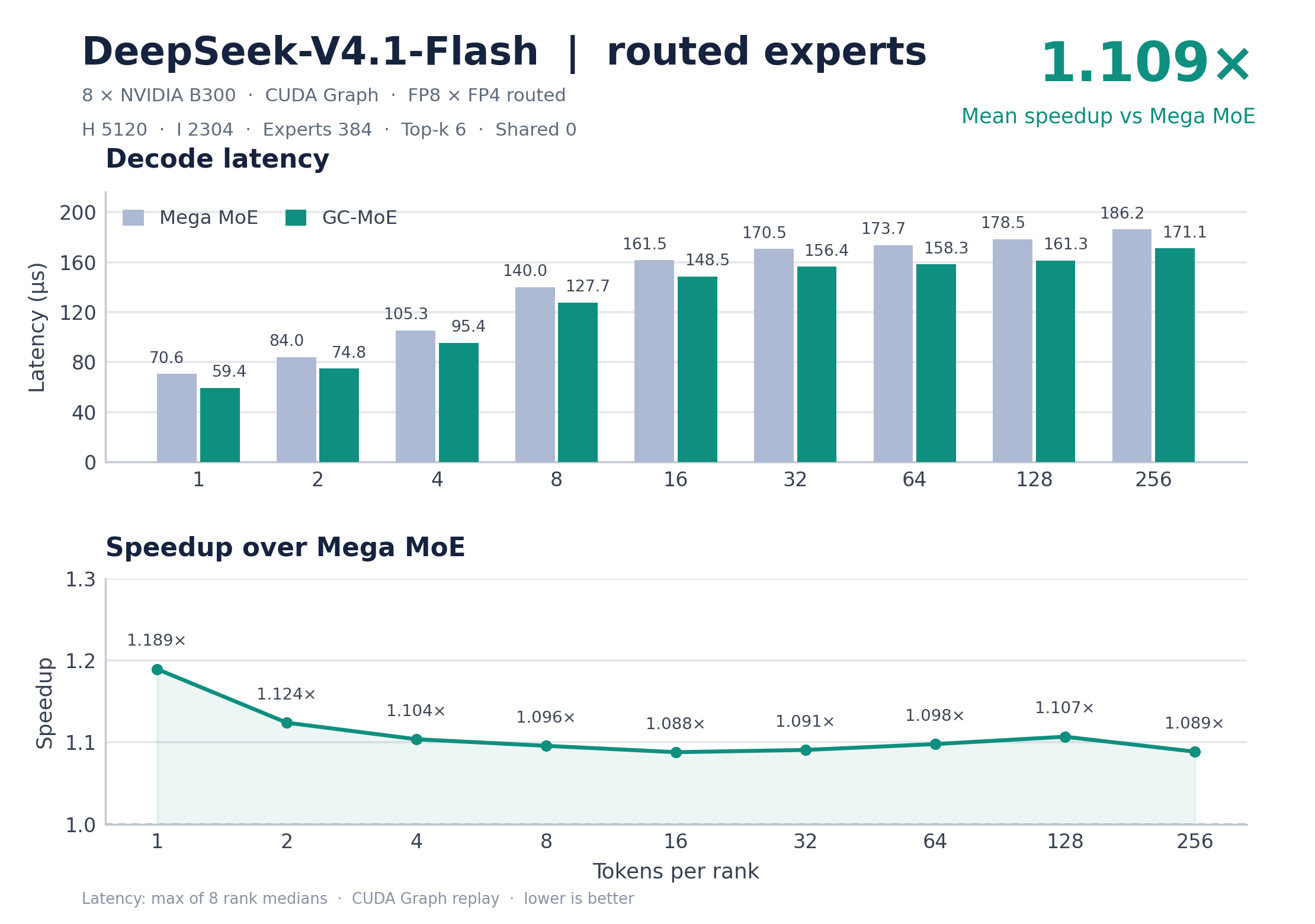}
\caption{DeepSeek-V4.1-Flash routed-expert decode on 8$\times$B300: latency and speedup.}
\label{fig:b300dsv}
\end{figure}

\paragraph{Blackwell (8$\times$B300).} Table~\ref{tab:b300} and Figures~\ref{fig:b300glm}--\ref{fig:b300dsv} report decode for both models. GC-MoE wins all 18 points. The compute partition has 128 SMs versus the baseline's 148, so the win is not from more arithmetic capacity: at decode the grouped GEMM is bandwidth-bound, and dedicated SMs plus the removed tail matter more than the 20 SMs ceded to dispatch. The mean speedup is 1.102$\times$ for GLM-5.2 and 1.109$\times$ for DeepSeek-V4.1-Flash (1.106$\times$ overall), and the per-point range is 1.085--1.189$\times$. The B300 curves are flatter than the H20 curve (1.085--1.189$\times$ versus a 1.26--1.30$\times$ plateau). For DeepSeek-V4.1-Flash the absolute saving stays within 9--18\,$\mu$s per layer across the sweep, so the relative gain is largest at 1 token per rank (1.189$\times$: 11.2\,$\mu$s on a 70\,$\mu$s layer) and flattens as the GEMM grows (1.089$\times$ at 256). For GLM-5.2 the saving instead grows with batch (5.6\,$\mu$s at 1 token/rank to 18.7\,$\mu$s at 256), consistent with the dedicated-SM GEMM advantage scaling with the amount of grouped-GEMM work rather than with the fixed tail (1.095$\times$ to 1.128$\times$). Per-rank medians are tightly clustered (at 1 token per rank for DeepSeek-V4.1-Flash, GC-MoE's eight rank medians span 59.21--59.39\,$\mu$s and the baseline's 70.22--70.63\,$\mu$s), so the speedup is not an artifact of one straggler rank.

\begin{table}[t]
\centering
\caption{Decode latency and speedup, 8$\times$B300 (routed experts), both models. Slowest-rank median per-invocation latency. DSV4.1-Flash abbreviates DeepSeek-V4.1-Flash.}
\label{tab:b300}
\small
\begin{tabular}{llrrrr}
\toprule
Model & Tokens/rank & Total tokens & Mega ($\mu$s) & GC-MoE ($\mu$s) & Speedup \\
\midrule
GLM-5.2 & 1 & 8 & 64.184 & 58.601 & 1.0953 \\
 & 2 & 16 & 87.457 & 79.228 & 1.1039 \\
 & 4 & 32 & 108.085 & 98.225 & 1.1004 \\
 & 8 & 64 & 125.241 & 115.041 & 1.0887 \\
 & 16 & 128 & 131.652 & 121.385 & 1.0846 \\
 & 32 & 256 & 133.215 & 121.493 & 1.0965 \\
 & 64 & 512 & 137.227 & 123.694 & 1.1094 \\
 & 128 & 1024 & 144.344 & 129.517 & 1.1145 \\
 & 256 & 2048 & 164.383 & 145.683 & 1.1284 \\
\midrule
DSV4.1-Flash & 1 & 8 & 70.630 & 59.387 & \best{1.1893} \\
 & 2 & 16 & 84.022 & 74.758 & 1.1239 \\
 & 4 & 32 & 105.265 & 95.373 & 1.1037 \\
 & 8 & 64 & 139.955 & 127.732 & 1.0957 \\
 & 16 & 128 & 161.549 & 148.495 & 1.0879 \\
 & 32 & 256 & 170.546 & 156.371 & 1.0907 \\
 & 64 & 512 & 173.747 & 158.264 & 1.0978 \\
 & 128 & 1024 & 178.495 & 161.271 & 1.1068 \\
 & 256 & 2048 & 186.200 & 171.051 & 1.0886 \\
\midrule
\multicolumn{5}{r}{Mean speedup: GLM-5.2 (9 points)} & \best{1.1024} \\
\multicolumn{5}{r}{Mean speedup: DeepSeek-V4.1-Flash (9 points)} & \best{1.1094} \\
\multicolumn{5}{r}{Mean speedup: overall (18 points)} & \best{1.1059} \\
\bottomrule
\end{tabular}
\end{table}

\subsection{Fused shared experts}
\label{sec:se-eval}

\begin{figure}[t]
\centering
\includegraphics[width=0.92\linewidth]{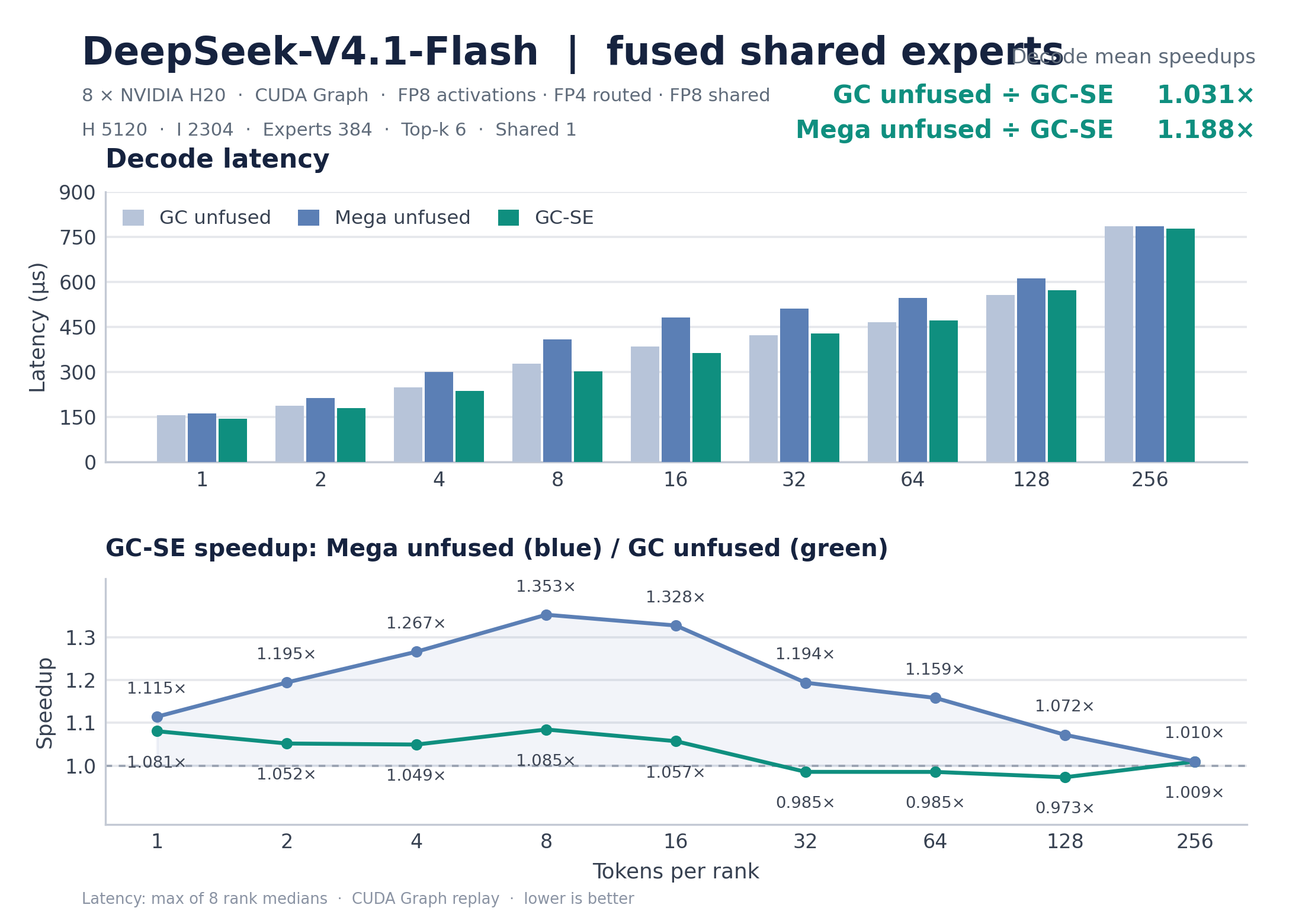}
\caption{DeepSeek-V4.1-Flash with shared experts on 8$\times$H20: gc-se versus gc-unfused and mega-unfused.}
\label{fig:h20se}
\end{figure}

\begin{figure}[t]
\centering
\includegraphics[width=0.92\linewidth]{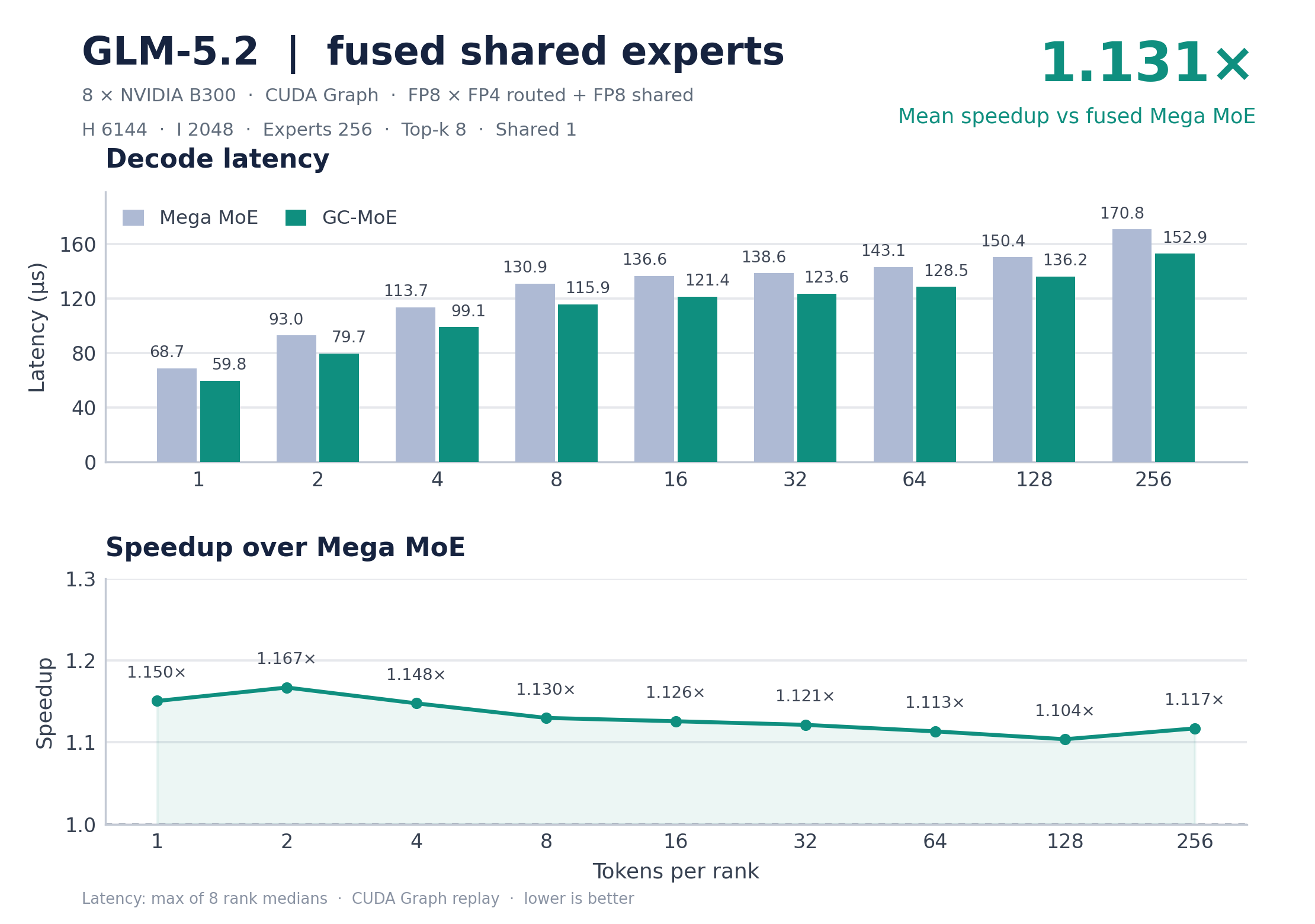}
\caption{GLM-5.2 with shared experts on 8$\times$B300: latency and speedup.}
\label{fig:b300se}
\end{figure}

\begin{figure}[t]
\centering
\includegraphics[width=0.92\linewidth]{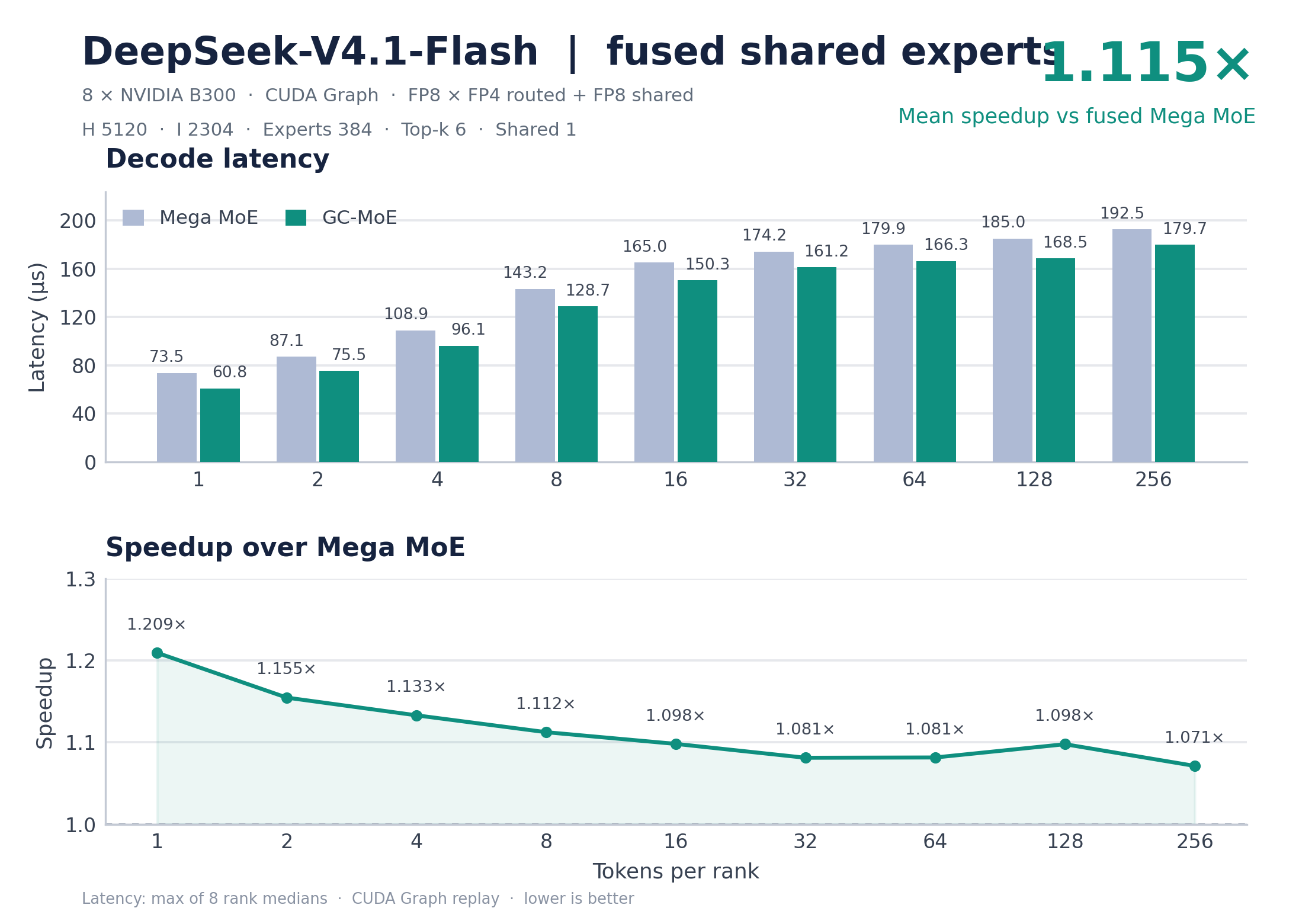}
\caption{DeepSeek-V4.1-Flash with shared experts on 8$\times$B300: latency and speedup.}
\label{fig:b300se2}
\end{figure}

Tables~\ref{tab:h20se} and~\ref{tab:b300se} and Figures~\ref{fig:h20se},~\ref{fig:b300se}, and~\ref{fig:b300se2} report the shared-expert setting. On H20, fusing one shared expert is 1.031$\times$ faster on average than GC-MoE without fusion (6 of 9 points), and 1.188$\times$ faster than the unfused Mega baseline (9 of 9 points), peaking at 1.353$\times$ at 8 tokens per rank. Because the H20 reference has no fused shared-expert path, its baseline is the routed Mega kernel plus a separate shared-expert kernel; the fused-versus-unfused comparison is the within-operator ablation. At small batch the shared expert is almost free because its GEMM interleaves with routed work on the compute partition and its output joins the existing combine reduction; at 32--128 tokens per rank it adds serial compute and the ratio dips slightly below one (0.973--0.985$\times$), which is the regime where the scheduler places shared tasks after routed work. On B300 the baseline runs its own fused shared-expert mode, so the comparison is between two feature-matched fused operators rather than a fusion ablation: GC-MoE is faster at all 18 points, with 1.131$\times$ mean for GLM-5.2 and 1.115$\times$ for DeepSeek-V4.1-Flash (1.123$\times$ overall) and a peak of 1.209$\times$ at 1 token per rank. Across the routed and shared-expert sweeps on both platforms, the evaluation covers 54 points (9 H20 routed, 9 H20 shared, 18 B300 routed, and 18 B300 shared), and GC-MoE is faster at every one.

\begin{table}[t]
\centering
\caption{Decode latency, fused shared experts, DeepSeek-V4.1-Flash on 8$\times$H20. \emph{gc-se}: GC-MoE with fusion; \emph{gc-unfused}: GC-MoE with the shared expert run separately; \emph{mega-unfused}: baseline. Ratios are against gc-se.}
\label{tab:h20se}
\small
\begin{tabular}{rrrrrr}
\toprule
Tokens/rank & gc-se ($\mu$s) & gc-unfused ($\mu$s) & unfused/se & mega-unfused ($\mu$s) & mega/se \\
\midrule
1 & 145.171 & 156.866 & 1.0806 & 161.804 & 1.1146 \\
2 & 179.286 & 188.566 & 1.0518 & 214.191 & 1.1947 \\
4 & 236.353 & 248.009 & 1.0493 & 299.391 & 1.2667 \\
8 & 302.667 & 328.262 & 1.0846 & 409.599 & \best{1.3533} \\
16 & 363.640 & 384.421 & 1.0571 & 482.938 & 1.3281 \\
32 & 428.722 & 422.456 & 0.9854 & 511.952 & 1.1941 \\
64 & 472.348 & 465.376 & 0.9852 & 547.238 & 1.1585 \\
128 & 572.091 & 556.513 & 0.9728 & 613.336 & 1.0721 \\
256 & 779.019 & 786.387 & 1.0095 & 786.631 & 1.0098 \\
\midrule
\multicolumn{3}{r}{Mean (9 points)} & \best{1.0307} & -- & \best{1.1880} \\
\bottomrule
\end{tabular}
\end{table}

\begin{table}[t]
\centering
\caption{Decode latency and speedup with fused shared experts, 8$\times$B300. The baseline is Mega MoE in its fused shared-expert mode. Slowest-rank median per-invocation latency. DSV4.1-Flash abbreviates DeepSeek-V4.1-Flash.}
\label{tab:b300se}
\small
\begin{tabular}{llrrrr}
\toprule
Model & Tokens/rank & Total tokens & Mega-se ($\mu$s) & GC-MoE ($\mu$s) & Speedup \\
\midrule
GLM-5.2 & 1 & 8 & 68.739 & 59.752 & 1.1504 \\
 & 2 & 16 & 93.033 & 79.730 & 1.1669 \\
 & 4 & 32 & 113.672 & 99.059 & 1.1475 \\
 & 8 & 64 & 130.904 & 115.870 & 1.1297 \\
 & 16 & 128 & 136.624 & 121.372 & 1.1257 \\
 & 32 & 256 & 138.641 & 123.645 & 1.1213 \\
 & 64 & 512 & 143.096 & 128.534 & 1.1133 \\
 & 128 & 1024 & 150.360 & 136.231 & 1.1037 \\
 & 256 & 2048 & 170.763 & 152.883 & 1.1170 \\
\midrule
DSV4.1-Flash & 1 & 8 & 73.541 & 60.816 & \best{1.2092} \\
 & 2 & 16 & 87.140 & 75.478 & 1.1545 \\
 & 4 & 32 & 108.908 & 96.132 & 1.1329 \\
 & 8 & 64 & 143.151 & 128.697 & 1.1123 \\
 & 16 & 128 & 164.977 & 150.251 & 1.0980 \\
 & 32 & 256 & 174.223 & 161.154 & 1.0811 \\
 & 64 & 512 & 179.868 & 166.320 & 1.0815 \\
 & 128 & 1024 & 184.979 & 168.524 & 1.0976 \\
 & 256 & 2048 & 192.473 & 179.688 & 1.0712 \\
\midrule
\multicolumn{5}{r}{Mean speedup: GLM-5.2 (9 points)} & \best{1.1306} \\
\multicolumn{5}{r}{Mean speedup: DeepSeek-V4.1-Flash (9 points)} & \best{1.1154} \\
\multicolumn{5}{r}{Mean speedup: overall (18 points)} & \best{1.1230} \\
\bottomrule
\end{tabular}
\end{table}

\subsection{Throughput and bandwidth analysis}
\label{sec:tput}

Tables~\ref{tab:h20tput} and~\ref{tab:b300tput} report effective throughput. Two conventions must be stated precisely. \emph{Effective TFLOPS} counts only matmul work ($6\times$assignments$\times H I$ for routed experts), divided by the slowest-rank latency and aggregated over eight GPUs. \emph{HBM working-set bandwidth} and \emph{NVLink payload bandwidth} are analytic models, not hardware counters: the former counts each owner's unique active expert weights and packed scales once, the latter a one-way per-assignment payload of $(H+H/32+2H)$ bytes, counting only assignments whose expert resides on a remote rank (an expected fraction $(P-1)/P$ of assignments).

The overlap raises effective arithmetic throughput at every point. On H20, GC-MoE reaches 95.6 versus 73.8 TFLOPS at 8 tokens per rank (+29.5\%, near the 1.296$\times$ peak speedup at 16) and 1287.8 versus 1251.4 at 256. On B300, GLM-5.2 reaches 8490.7 versus 7524.8 TFLOPS at 256 tokens per rank (+12.8\%), and at 1 token per rank even the small-batch case improves from 75.3 to 82.5 TFLOPS. The working-set bandwidth model shows the same picture: H20 peaks at 18\,931.9 versus 14\,973.9 GB/s at 32 tokens per rank (+26.4\%). The NVLink payload is identical by construction, and the modeled payload bandwidth rises with the speedup (239.7 to 246.7 GB/s at the largest H20 batch, +2.9\%): the same bytes move in less wall time because dispatch no longer competes with the GEMM for SM resources and each replay avoids the trailing cross-rank barrier, so the partition does not cost cross-rank bandwidth.

\begin{table}[t]
\centering
\caption{Effective throughput and analytic bandwidth models, DeepSeek-V4.1-Flash on 8$\times$H20. TFLOPS counts matmul work only; HBM is the expert weight+scale working-set model; NVLink is the one-way assignment payload model. All values are aggregate over eight GPUs; GC denotes GC-MoE.}
\label{tab:h20tput}
\small
\begin{tabular}{rrrrrrr}
\toprule
& \multicolumn{2}{c}{TFLOPS} & \multicolumn{2}{c}{HBM (GB/s)} & \multicolumn{2}{c}{NVLink (GB/s)} \\
\cmidrule(lr){2-3}\cmidrule(lr){4-5}\cmidrule(lr){6-7}
Tokens/rank & Mega & GC & Mega & GC & Mega & GC \\
\midrule
1 & 28.297 & 29.501 & 7\,516.4 & 7\,836.2 & 5.171 & 5.391 \\
2 & 40.125 & 47.101 & 9\,770.0 & 11\,468.7 & 7.424 & 8.714 \\
4 & 53.080 & 67.079 & 11\,235.5 & 14\,198.7 & 9.942 & 12.564 \\
8 & 73.805 & 95.597 & 12\,814.4 & 16\,598.0 & 14.076 & 18.233 \\
16 & 122.782 & 159.087 & 14\,311.1 & 18\,542.7 & 23.522 & 30.478 \\
32 & 230.287 & 291.158 & 14\,973.9 & 18\,931.9 & 44.743 & 56.570 \\
64 & 428.686 & 502.870 & 14\,233.7 & 16\,696.9 & 81.944 & 96.124 \\
128 & 763.452 & 881.733 & 12\,674.5 & 14\,638.2 & 147.052 & 169.835 \\
256 & 1\,251.393 & 1\,287.785 & 10\,387.5 & 10\,689.6 & 239.719 & 246.690 \\
\bottomrule
\end{tabular}
\end{table}

\begin{table}[t]
\centering
\caption{Effective TFLOPS (matmul work only, aggregate over eight GPUs), 8$\times$B300, routed experts. GC denotes GC-MoE.}
\label{tab:b300tput}
\small
\begin{tabular}{rrrrr}
\toprule
& \multicolumn{2}{c}{GLM-5.2} & \multicolumn{2}{c}{DeepSeek-V4.1-Flash} \\
\cmidrule(lr){2-3}\cmidrule(lr){4-5}
Tokens/rank & Mega & GC & Mega & GC \\
\midrule
1 & 75.280 & 82.453 & 48.101 & 57.208 \\
2 & 110.496 & 121.974 & 80.869 & 90.891 \\
4 & 178.817 & 196.765 & 129.099 & 142.488 \\
8 & 308.642 & 336.007 & 194.199 & 212.781 \\
16 & 587.226 & 636.895 & 336.480 & 366.062 \\
32 & 1\,160.668 & 1\,272.653 & 637.460 & 695.245 \\
64 & 2\,253.475 & 2\,500.026 & 1\,251.436 & 1\,373.857 \\
128 & 4\,284.727 & 4\,775.234 & 2\,436.293 & 2\,696.491 \\
256 & 7\,524.828 & 8\,490.679 & 4\,670.942 & 5\,084.628 \\
\bottomrule
\end{tabular}
\end{table}

\subsection{Numerical accuracy}
\label{sec:accuracy}

GC-MoE uses the caller's quantized inputs and the baseline's quantized weights and performs the same low-precision arithmetic; the difference is scheduling, layout, and instruction selection. Table~\ref{tab:acc} reports the H20 routed-expert accuracy suite: 47 cases (9 default, 23 boundary, 15 asymmetric, including empty ranks) over all eight ranks. All 376 rank-cases pass the cosine thresholds (0.97 against BF16 for GC-MoE and the baseline, 0.97 for GC-versus-Mega), and the processed-difference and graph/eager repeat checks are within 0.002 (repeat difference 0). The processed difference is $1-2\langle y,\hat y\rangle/(\|y\|^2+\|\hat y\|^2)$ against a reference forward that dequantizes the packed weights and evaluates the layer in FP32. GC-MoE's minimum cosine against the baseline is 0.9999998, and the two operators have identical maximum absolute error (0.109375) and identical maximum relative L2 error (0.00600836286321): at BF16 output precision they share the same maximum-error envelope, while their outputs still differ slightly. The H20 shared-expert suite likewise passes all 47 cases and 376 rank checks (cosine $\geq0.99$), with a minimum GC-versus-Mega cosine of 0.9999997 (maximum absolute error 0.0625, maximum relative L2 $5.81\times10^{-5}$). On B300, all 94 cases and 752 rank checks pass with cosine $\geq0.97$ (routed) and $\geq0.99$ (shared experts), with all finite checks passing. The shared-expert path stages scales in the same permuted order as the routed path, and the fused reduction accumulates BF16 values into FP32 in top-$k$ slot order, which is why fusion does not move accuracy.

\begin{table}[t]
\centering
\caption{Numerical accuracy, DeepSeek-V4.1-Flash on 8$\times$H20, routed experts. Thresholds: cosine $\geq0.97$; processed diff and eager/graph repeat diff $\leq0.002$. Cosine columns are per-category minima.}
\label{tab:acc}
\small
\begin{tabular}{lrrrrr}
\toprule
Category & Cases & GC/BF16 & Mega/BF16 & GC/Mega & Max proc. diff \\
\midrule
default & 9/9 & 0.977039 & 0.977039 & 0.99999976 & $1.80\times10^{-5}$ \\
boundary & 23/23 & 0.977672 & 0.977672 & 0.99999976 & $9.95\times10^{-6}$ \\
asymmetric & 15/15 & 0.977837 & 0.977837 & 0.99999976 & $1.80\times10^{-5}$ \\
\midrule
\multicolumn{6}{l}{GC: max absolute error 0.109375; max relative L2 0.00600836286321; repeat diff 0.} \\
\multicolumn{6}{l}{Mega: max absolute error 0.109375; max relative L2 0.00600836286321.} \\
\bottomrule
\end{tabular}
\end{table}

\clearpage
\section{Related Work}
\label{sec:related}

\paragraph{MoE models and systems.} Sparsely gated MoE introduced conditional computation with top-$k$ routing~\citep{shazeer2017}; GShard and Switch Transformers established expert parallelism and all-to-all dispatch at scale~\citep{gshard,switch}; recent models use fine-grained experts and low precision~\citep{mixtral,deepseekv2,deepseekv3}. Systems work optimizes MoE training and inference at the framework level: DeepEP provides low-latency all-to-all kernels for expert parallelism~\citep{deepep}; MegaBlocks casts MoE as block-sparse GEMM with no token dropping~\citep{megablocks}; Tutel adapts parallelism and pipelining~\citep{tutel}; FasterMoE models and overlaps communication with computation across experts~\citep{fastermoe}; Lina's non-blocking all-to-all overlaps communication with expert computation across batches~\citep{lina}; SonicMoE optimizes MoE kernels for high sparsity and IO~\citep{sonicmoe}. These systems overlap communication with compute \emph{across} units of work (micro-batches, layers, experts). GC-MoE instead places dispatch and compute on disjoint SM sets \emph{inside one layer}, so the overlap costs the compute SMs no dispatch warps or staging buffers; this is orthogonal and composable with the above. Beyond overlap, the decomposition gives each phase its own resource and synchronization budget, so dispatch and GEMM can be tuned independently; this matters most when the layer is bandwidth-bound, as in decode.

\paragraph{Efficient kernels and serving.} FlashAttention established IO-aware tiling~\citep{fa1}, its successors exploit parallelism and Hopper asynchrony with FP8~\citep{fa2,fa3}; Triton and CUTLASS provide the tiled-kernel substrate~\citep{triton,cutlass}. DeepGEMM provides the FP8/FP4 grouped-GEMM family, including the Mega MoE baseline, which already overlaps dispatch with GEMM by sharing every SM; GC-MoE restructures that shared-SM, warp-specialized organization into disjoint SM partitions~\citep{deepgemm}; FlashInfer contributes inference kernels~\citep{flashinfer}. Serving engines schedule requests and memory~\citep{vllm,sarathi}; GC-MoE is an operator that such engines can call under CUDA Graph capture.

\paragraph{Low-precision formats.} FP8 formats and their scaling behavior were characterized early~\citep{fp8quant}, and FP8 training/inference stacks established per-group scaling~\citep{fp8lm,deepseekv3}; NVFP4 and MXFP4 define the FP4 microscaling conventions used here~\citep{nvfp4,mxfp4}. GC-MoE's contribution is not a new format but the dual-architecture realization of FP8$\times$FP4 grouped GEMM: hardware block scaling with K128 weight prepacking and partition-level scheduling on Blackwell, and in-register FP4 decoding with window-based scale repacking on Hopper (Section~\ref{sec:gemm}).

\paragraph{GPU partitioning and execution control.} CUDA Green Contexts are a driver-level SM-partitioning primitive~\citep{cuda}; MPS and MIG provide coarser isolation. Prior work uses green contexts and spatial partitioning for \emph{isolation}: characterizing green-context, MPS, and MIG isolation~\citep{gcisolation}, co-locating models on spatially partitioned GPUs~\citep{eclip}, and isolating prefill/decode slots in an agentic serving system~\citep{agentserve}. To our knowledge, GC-MoE is the first operator to use green contexts for intra-operator communication--computation overlap, with co-resident kernels synchronized at ring-block granularity.

\section{Conclusion}
\label{sec:conclusion}

GC-MoE organizes the expert-parallel MoE layer as a three-node CUDA Graph (four on Hopper with fused shared experts) across two CUDA Green Contexts: a small communication partition streams tokens into a ring while a large compute partition consumes ring blocks with FP8$\times$FP4 tensor-core work, and a finalize node reduces the outputs without a trailing cross-rank barrier. At decode the grouped GEMM is bandwidth-bound, so the 128-SM compute partition loses little by ceding 20 SMs to dispatch, while each role gains its own resource and synchronization budget and can be optimized independently. The design required three ingredients beyond the partition itself---generation-tagged ring synchronization, double-buffered cross-rank counters, and architecture-specific low-precision MMA paths for Blackwell and Hopper---and it yields 1.188$\times$ mean decode speedup on 8$\times$H20 and 1.106--1.123$\times$ on 8$\times$B300 across two production MoE configurations, with accuracy that matches the baseline's error envelope at BF16 precision (minimum cosine 0.9999997 against the baseline on both measured H20 suites).

\paragraph{Limitations.} The operator targets decode; prefill-sized token counts are supported but the compute partition, not overlap, becomes the binding resource, and at the largest measured batch (256 tokens per rank) the gain narrows to 1.01--1.13$\times$. The shape constraints of Appendix~\ref{app:constraints} reflect hard design decisions (64-bit completion masks, static signal sizing, a 72-rank by-value peer table). A controlled ablation of partition size and an overlap-on/off comparison are not included in this version; the fused-versus-unfused shared-expert comparison is the only within-operator ablation reported. The workspace is single-instance, so calls that share it must be serialized.

\paragraph{Future work.} We plan controlled experiments over partition sizes and dispatch SM counts, extension to prefill-heavy regimes and additional model families, adaptive partitioning when the driver permits finer splits, and integration into a graph-capturing serving stack.

\bibliographystyle{plainnat}
\bibliography{refs}
\clearpage
\appendix

\section{Operator constraints}
\label{app:constraints}

Table~\ref{tab:constraints} lists the supported envelope and the technical reason for each constraint.

\begin{table}[ht]
\centering
\caption{Supported envelope and rationale (SM103 path unless noted).}
\label{tab:constraints}
\small
\begin{tabular}{L{0.30\linewidth}L{0.64\linewidth}}
\toprule
Constraint & Technical reason \\
\midrule
$I\leq 4096$, $I \bmod 128 = 0$ & W1 output tiles $2I/128$ must fit the 64-bit \texttt{l2\_full\_mask}; K128 tiles complete. \\
$H \bmod 256 = 0$ & 2-CTA clusters pair N128 tiles, requiring $H/128$ even. \\
$k + \mathbf{1}[\text{shared}]\leq 32$, $k\geq1$ & One warp lane carries one combine slot; dispatch reads $32/k$ tokens per warp. \\
$E\leq2048$ global, $\leq512$ per rank, $E \bmod P=0$ & Static per-expert signal region; lane-held per-expert counters. \\
$P\leq72$ ranks & 592-byte by-value peer descriptor with 72 offsets. \\
Routed recipe $(1,1,32)$; SwiGLU & MMA scale-factor path hard-codes K32 UE8M0; SwiGLU fused in the W1 epilogue. \\
$B_M\in\{16,32,64,96,128,192,240\}$ & 2-CTA MMA alignment; TMEM/SMEM budget. \\
Shared recipe $(1,1,32)$ or $(1,128,128)$ & Scale staging and selector logic implemented for these K granularities. \\
One process per GPU, peer-accessible & Symmetric memory, green contexts, and captured graph nodes. \\
Serialized calls per workspace & Persistent single-instance ring and signal state. \\
\bottomrule
\end{tabular}
\end{table}

\end{document}